\documentclass[%
 reprint,
superscriptaddress,
preprintnumbers,
nofootinbib,
 amsmath,amssymb,
 aps,
prd,
]{revtex4-2}

\usepackage{graphicx}
\usepackage{dcolumn}
\usepackage{bm}
\usepackage{aas_macros}
\usepackage{hyperref}

\newcommand{\redmapper}{redMaPPer}

\newcommand{\msun}{M_\odot}
\newcommand{\msunh}{M_{\odot}h^{-1}}
\newcommand{\vel}{\,{\rm km\,s^{-1}}}
\newcommand{\de}{\text{d}}

\newcommand{\Mpc}{\rm{Mpc}}

\newcommand{\LCDM}{$\Lambda$CDM }

\newcommand{\ob}{^{\rm ob}}

\newcommand{\zmax}{z_{\rm max}}

\newcommand{\lob}{\lambda^{\rm ob}}
\newcommand{\ltrue}{\lambda^{\rm true}}

\newcommand{\mwl}{M_{\rm WL}}
\newcommand{\sz}{{\rm SZ}}

\usepackage[usenames]{color}
\definecolor{purple}{RGB}{150,0,200}

\newcommand{\planck}{{\it Planck}}

\newcommand{\hMpc}{h^{-1}\ \Mpc}

\newcommand{\hatn}{\boldsymbol{\hat{n}}}
\newcommand{\hinv}{h^{-1}}
\newcommand{\Lmin}{L_{\rm min}}
\newcommand{\fcen}{f_{\rm cen}}

\newcommand{\Omegam}{\Omega_{{\rm m}}}
\newcommand{\Omegab}{\Omega_{{\rm b}}}

\newcommand{\Planck}{{\it Planck}}

\defcitealias{desy1wl}{MV19}
\defcitealias{Bocquet2018}{B19}
\defcitealias{desy1cl}{DES20}

\begin{document}

\preprint{DES-2020-0588}
\preprint{FERMILAB-PUB-20-541-V}


\title[DES Y1 x SPT Cluster Cosmology]{Cosmological Constraints from DES Y1 Cluster Abundances and SPT Multi-wavelength data}


\author{M.~Costanzi}
\affiliation{INAF-Osservatorio Astronomico di Trieste, via G. B. Tiepolo 11, I-34143 Trieste, Italy}
\affiliation{Astronomy Unit, Department of Physics, University of Trieste, via Tiepolo 11, I-34131 Trieste, Italy}
\affiliation{Institute for Fundamental Physics of the Universe, Via Beirut 2, 34014 Trieste, Italy}
\author{A.~Saro}
\affiliation{INAF-Osservatorio Astronomico di Trieste, via G. B. Tiepolo 11, I-34143 Trieste, Italy}
\affiliation{Astronomy Unit, Department of Physics, University of Trieste, via Tiepolo 11, I-34131 Trieste, Italy}
\affiliation{Institute for Fundamental Physics of the Universe, Via Beirut 2, 34014 Trieste, Italy}
\affiliation{INFN - National Institute for Nuclear Physics, Via Valerio 2, I-34127 Trieste, Italy}
\author{S.~Bocquet}
\affiliation{Faculty of Physics, Ludwig-Maximilians-Universit\"at, Scheinerstr. 1, 81679 Munich, Germany}
\author{T.~M.~C.~Abbott}
\affiliation{Cerro Tololo Inter-American Observatory, NSF's National Optical-Infrared Astronomy Research Laboratory, Casilla 603, La Serena, Chile}
\author{M.~Aguena}
\affiliation{Departamento de F\'isica Matem\'atica, Instituto de F\'isica, Universidade de S\~ao Paulo, CP 66318, S\~ao Paulo, SP, 05314-970, Brazil}
\affiliation{Laborat\'orio Interinstitucional de e-Astronomia - LIneA, Rua Gal. Jos\'e Cristino 77, Rio de Janeiro, RJ - 20921-400, Brazil}
\author{S.~Allam}
\affiliation{Fermi National Accelerator Laboratory, P. O. Box 500, Batavia, IL 60510, USA}
\author{A.~Amara}
\affiliation{Institute of Cosmology and Gravitation, University of Portsmouth, Portsmouth, PO1 3FX, UK}
\author{J.~Annis}
\affiliation{Fermi National Accelerator Laboratory, P. O. Box 500, Batavia, IL 60510, USA}
\author{S.~Avila}
\affiliation{Instituto de Fisica Teorica UAM/CSIC, Universidad Autonoma de Madrid, 28049 Madrid, Spain}
\author{D.~Bacon}
\affiliation{Institute of Cosmology and Gravitation, University of Portsmouth, Portsmouth, PO1 3FX, UK}
\author{B.~A.~Benson}
\affiliation{Fermi National Accelerator Laboratory, Batavia, IL 60510-0500, USA}
\affiliation{Department of Astronomy and Astrophysics, University of Chicago, 5640 South Ellis Avenue, Chicago, IL 60637}
\affiliation{Kavli Institute for Cosmological Physics, University of Chicago, 5640 South Ellis Avenue, Chicago, IL 60637}
\author{S.~Bhargava}
\affiliation{Department of Physics and Astronomy, Pevensey Building, University of Sussex, Brighton, BN1 9QH, UK}
\author{D.~Brooks}
\affiliation{Department of Physics \& Astronomy, University College London, Gower Street, London, WC1E 6BT, UK}
\author{E.~Buckley-Geer}
\affiliation{Fermi National Accelerator Laboratory, P. O. Box 500, Batavia, IL 60510, USA}
\author{D.~L.~Burke}
\affiliation{Kavli Institute for Particle Astrophysics \& Cosmology, P. O. Box 2450, Stanford University, Stanford, CA 94305, USA}
\affiliation{SLAC National Accelerator Laboratory, Menlo Park, CA 94025, USA}
\author{A.~Carnero~Rosell}
\affiliation{Instituto de Astrofisica de Canarias, E-38205 La Laguna, Tenerife, Spain}
\affiliation{Universidad de La Laguna, Dpto. Astrofisica, E-38206 La Laguna, Tenerife, Spain}
\author{M.~Carrasco~Kind}
\affiliation{Department of Astronomy, University of Illinois at Urbana-Champaign, 1002 W. Green Street, Urbana, IL 61801, USA}
\affiliation{National Center for Supercomputing Applications, 1205 West Clark St., Urbana, IL 61801, USA}
\author{J.~Carretero}
\affiliation{Institut de F\'{\i}sica d'Altes Energies (IFAE), The Barcelona Institute of Science and Technology, Campus UAB, 08193 Bellaterra (Barcelona) Spain}
\author{A.~Choi}
\affiliation{Center for Cosmology and Astro-Particle Physics, The Ohio State University, Columbus, OH 43210, USA}
\author{L.~N.~da Costa}
\affiliation{Laborat\'orio Interinstitucional de e-Astronomia - LIneA, Rua Gal. Jos\'e Cristino 77, Rio de Janeiro, RJ - 20921-400, Brazil}
\affiliation{Observat\'orio Nacional, Rua Gal. Jos\'e Cristino 77, Rio de Janeiro, RJ - 20921-400, Brazil}
\author{M.~E.~S.~Pereira}
\affiliation{Department of Physics, University of Michigan, Ann Arbor, MI 48109, USA}
\author{J.~De~Vicente}
\affiliation{Centro de Investigaciones Energ\'eticas, Medioambientales y Tecnol\'ogicas (CIEMAT), Madrid, Spain}
\author{S.~Desai}
\affiliation{Department of Physics, IIT Hyderabad, Kandi, Telangana 502285, India}
\author{H.~T.~Diehl}
\affiliation{Fermi National Accelerator Laboratory, P. O. Box 500, Batavia, IL 60510, USA}
\author{J.~P.~Dietrich}
\affiliation{Faculty of Physics, Ludwig-Maximilians-Universit\"at, Scheinerstr. 1, 81679 Munich, Germany}
\author{P.~Doel}
\affiliation{Department of Physics \& Astronomy, University College London, Gower Street, London, WC1E 6BT, UK}
\author{T.~F.~Eifler}
\affiliation{Department of Astronomy/Steward Observatory, University of Arizona, 933 North Cherry Avenue, Tucson, AZ 85721-0065, USA}
\affiliation{Jet Propulsion Laboratory, California Institute of Technology, 4800 Oak Grove Dr., Pasadena, CA 91109, USA}
\author{S.~Everett}
\affiliation{Santa Cruz Institute for Particle Physics, Santa Cruz, CA 95064, USA}
\author{I.~Ferrero}
\affiliation{Institute of Theoretical Astrophysics, University of Oslo. P.O. Box 1029 Blindern, NO-0315 Oslo, Norway}
\author{A.~Fert\'e}
\affiliation{Jet Propulsion Laboratory, California Institute of Technology, 4800 Oak Grove Dr., Pasadena, CA 91109, USA}
\author{B.~Flaugher}
\affiliation{Fermi National Accelerator Laboratory, P. O. Box 500, Batavia, IL 60510, USA}
\author{P.~Fosalba}
\affiliation{Institut d'Estudis Espacials de Catalunya (IEEC), 08034 Barcelona, Spain}
\affiliation{Institute of Space Sciences (ICE, CSIC),  Campus UAB, Carrer de Can Magrans, s/n,  08193 Barcelona, Spain}
\author{J.~Frieman}
\affiliation{Fermi National Accelerator Laboratory, P. O. Box 500, Batavia, IL 60510, USA}
\affiliation{Kavli Institute for Cosmological Physics, University of Chicago, Chicago, IL 60637, USA}
\author{J.~Garc\'ia-Bellido}
\affiliation{Instituto de Fisica Teorica UAM/CSIC, Universidad Autonoma de Madrid, 28049 Madrid, Spain}
\author{E.~Gaztanaga}
\affiliation{Institut d'Estudis Espacials de Catalunya (IEEC), 08034 Barcelona, Spain}
\affiliation{Institute of Space Sciences (ICE, CSIC),  Campus UAB, Carrer de Can Magrans, s/n,  08193 Barcelona, Spain}
\author{D.~W.~Gerdes}
\affiliation{Department of Astronomy, University of Michigan, Ann Arbor, MI 48109, USA}
\affiliation{Department of Physics, University of Michigan, Ann Arbor, MI 48109, USA}
\author{T.~Giannantonio}
\affiliation{Institute of Astronomy, University of Cambridge, Madingley Road, Cambridge CB3 0HA, UK}
\affiliation{Kavli Institute for Cosmology, University of Cambridge, Madingley Road, Cambridge CB3 0HA, UK}
\author{P.~Giles}
\affiliation{Department of Physics and Astronomy, Pevensey Building, University of Sussex, Brighton, BN1 9QH, UK}
\author{S.~Grandis}
\affiliation{Faculty of Physics, Ludwig-Maximilians-Universit\"at, Scheinerstr. 1, 81679 Munich, Germany}
\author{D.~Gruen}
\affiliation{Department of Physics, Stanford University, 382 Via Pueblo Mall, Stanford, CA 94305, USA}
\affiliation{Kavli Institute for Particle Astrophysics \& Cosmology, P. O. Box 2450, Stanford University, Stanford, CA 94305, USA}
\affiliation{SLAC National Accelerator Laboratory, Menlo Park, CA 94025, USA}
\author{R.~A.~Gruendl}
\affiliation{Department of Astronomy, University of Illinois at Urbana-Champaign, 1002 W. Green Street, Urbana, IL 61801, USA}
\affiliation{National Center for Supercomputing Applications, 1205 West Clark St., Urbana, IL 61801, USA}
\author{N.~Gupta}
\affiliation{School of Physics, University of Melbourne, Parkville, VIC 3010, Australia}
\author{G.~Gutierrez}
\affiliation{Fermi National Accelerator Laboratory, P. O. Box 500, Batavia, IL 60510, USA}
\author{W.~G.~Hartley}
\affiliation{D\'{e}partement de Physique Th\'{e}orique and Center for Astroparticle Physics, Universit\'{e} de Gen\`{e}ve, 24 quai Ernest Ansermet, CH-1211 Geneva, Switzerland}
\author{S.~R.~Hinton}
\affiliation{School of Mathematics and Physics, University of Queensland,  Brisbane, QLD 4072, Australia}
\author{D.~L.~Hollowood}
\affiliation{Santa Cruz Institute for Particle Physics, Santa Cruz, CA 95064, USA}
\author{K.~Honscheid}
\affiliation{Center for Cosmology and Astro-Particle Physics, The Ohio State University, Columbus, OH 43210, USA}
\affiliation{Department of Physics, The Ohio State University, Columbus, OH 43210, USA}
\author{D.~J.~James}
\affiliation{Center for Astrophysics $\vert$ Harvard \& Smithsonian, 60 Garden Street, Cambridge, MA 02138, USA}
\author{T.~Jeltema}
\affiliation{Santa Cruz Institute for Particle Physics, Santa Cruz, CA 95064, USA}
\author{E.~Krause}
\affiliation{Department of Astronomy/Steward Observatory, University of Arizona, 933 North Cherry Avenue, Tucson, AZ 85721-0065, USA}
\author{K.~Kuehn}
\affiliation{Australian Astronomical Optics, Macquarie University, North Ryde, NSW 2113, Australia}
\affiliation{Lowell Observatory, 1400 Mars Hill Rd, Flagstaff, AZ 86001, USA}
\author{N.~Kuropatkin}
\affiliation{Fermi National Accelerator Laboratory, P. O. Box 500, Batavia, IL 60510, USA}
\author{O.~Lahav}
\affiliation{Department of Physics \& Astronomy, University College London, Gower Street, London, WC1E 6BT, UK}
\author{M.~Lima}
\affiliation{Departamento de F\'isica Matem\'atica, Instituto de F\'isica, Universidade de S\~ao Paulo, CP 66318, S\~ao Paulo, SP, 05314-970, Brazil}
\affiliation{Laborat\'orio Interinstitucional de e-Astronomia - LIneA, Rua Gal. Jos\'e Cristino 77, Rio de Janeiro, RJ - 20921-400, Brazil}
\author{N.~MacCrann}
\affiliation{Center for Cosmology and Astro-Particle Physics, The Ohio State University, Columbus, OH 43210, USA}
\affiliation{Department of Physics, The Ohio State University, Columbus, OH 43210, USA}
\author{M.~A.~G.~Maia}
\affiliation{Laborat\'orio Interinstitucional de e-Astronomia - LIneA, Rua Gal. Jos\'e Cristino 77, Rio de Janeiro, RJ - 20921-400, Brazil}
\affiliation{Observat\'orio Nacional, Rua Gal. Jos\'e Cristino 77, Rio de Janeiro, RJ - 20921-400, Brazil}
\author{J.~L.~Marshall}
\affiliation{George P. and Cynthia Woods Mitchell Institute for Fundamental Physics and Astronomy, and Department of Physics and Astronomy, Texas A\&M University, College Station, TX 77843,  USA}
\author{F.~Menanteau}
\affiliation{Department of Astronomy, University of Illinois at Urbana-Champaign, 1002 W. Green Street, Urbana, IL 61801, USA}
\affiliation{National Center for Supercomputing Applications, 1205 West Clark St., Urbana, IL 61801, USA}
\author{R.~Miquel}
\affiliation{Instituci\'o Catalana de Recerca i Estudis Avan\c{c}ats, E-08010 Barcelona, Spain}
\affiliation{Institut de F\'{\i}sica d'Altes Energies (IFAE), The Barcelona Institute of Science and Technology, Campus UAB, 08193 Bellaterra (Barcelona) Spain}
\author{J.~J.~Mohr}
\affiliation{Faculty of Physics, Ludwig-Maximilians-Universit\"at, Scheinerstr. 1, 81679 Munich, Germany}
\affiliation{Max Planck Institute for Extraterrestrial Physics, Giessenbachstrasse, 85748 Garching, Germany}
\author{R.~Morgan}
\affiliation{Physics Department, 2320 Chamberlin Hall, University of Wisconsin-Madison, 1150 University Avenue Madison, WI  53706-1390}
\author{J.~Myles}
\affiliation{Department of Physics, Stanford University, 382 Via Pueblo Mall, Stanford, CA 94305, USA}
\author{R.~L.~C.~Ogando}
\affiliation{Laborat\'orio Interinstitucional de e-Astronomia - LIneA, Rua Gal. Jos\'e Cristino 77, Rio de Janeiro, RJ - 20921-400, Brazil}
\affiliation{Observat\'orio Nacional, Rua Gal. Jos\'e Cristino 77, Rio de Janeiro, RJ - 20921-400, Brazil}
\author{A.~Palmese}
\affiliation{Fermi National Accelerator Laboratory, P. O. Box 500, Batavia, IL 60510, USA}
\affiliation{Kavli Institute for Cosmological Physics, University of Chicago, Chicago, IL 60637, USA}
\author{F.~Paz-Chinch\'{o}n}
\affiliation{Institute of Astronomy, University of Cambridge, Madingley Road, Cambridge CB3 0HA, UK}
\affiliation{National Center for Supercomputing Applications, 1205 West Clark St., Urbana, IL 61801, USA}
\author{A.~A.~Plazas}
\affiliation{Department of Astrophysical Sciences, Princeton University, Peyton Hall, Princeton, NJ 08544, USA}
\author{D.~Rapetti}
\affiliation{NASA Ames Research Center, Moffett Field, CA 94035, USA}
\affiliation{NASA Academic Mission Services Task Lead, Research Institute for Advanced Computer Science, Universities Space Research Association, Mountain View, CA 94043, USA}
\affiliation{Center for Astrophysics and Space Astronomy, Department of Astrophysical and Planetary Sciences, University of Colorado Boulder,CO 80309, USA}
\author{C.~L.~Reichardt}
\affiliation{School of Physics, University of Melbourne, Parkville, VIC 3010, Australia}
\author{A.~K.~Romer}
\affiliation{Department of Physics and Astronomy, Pevensey Building, University of Sussex, Brighton, BN1 9QH, UK}
\author{A.~Roodman}
\affiliation{Kavli Institute for Particle Astrophysics \& Cosmology, P. O. Box 2450, Stanford University, Stanford, CA 94305, USA}
\affiliation{SLAC National Accelerator Laboratory, Menlo Park, CA 94025, USA}
\author{F.~Ruppin}
\affiliation{Kavli Institute for Astrophysics and Space Research, Massachusetts Institute of Technology, Cambridge, MA 02139, USA}
\author{L.~Salvati}
\affiliation{INAF-Osservatorio Astronomico di Trieste, via G. B. Tiepolo 11, I-34143 Trieste, Italy}
\affiliation{Institute for Fundamental Physics of the Universe, Via Beirut 2, 34014 Trieste, Italy}
\author{S.~Samuroff}
\affiliation{Department of Physics, Carnegie Mellon University, Pittsburgh, Pennsylvania 15312, USA}
\author{E.~Sanchez}
\affiliation{Centro de Investigaciones Energ\'eticas, Medioambientales y Tecnol\'ogicas (CIEMAT), Madrid, Spain}
\author{V.~Scarpine}
\affiliation{Fermi National Accelerator Laboratory, P. O. Box 500, Batavia, IL 60510, USA}
\author{S.~Serrano}
\affiliation{Institut d'Estudis Espacials de Catalunya (IEEC), 08034 Barcelona, Spain}
\affiliation{Institute of Space Sciences (ICE, CSIC),  Campus UAB, Carrer de Can Magrans, s/n,  08193 Barcelona, Spain}
\author{I.~Sevilla-Noarbe}
\affiliation{Centro de Investigaciones Energ\'eticas, Medioambientales y Tecnol\'ogicas (CIEMAT), Madrid, Spain}
\author{P.~Singh}
\affiliation{INAF-Osservatorio Astronomico di Trieste, via G. B. Tiepolo 11, I-34143 Trieste, Italy}
\affiliation{Institute for Fundamental Physics of the Universe, Via Beirut 2, 34014 Trieste, Italy}
\author{M.~Smith}
\affiliation{School of Physics and Astronomy, University of Southampton,  Southampton, SO17 1BJ, UK}
\author{M.~Soares-Santos}
\affiliation{Department of Physics, University of Michigan, Ann Arbor, MI 48109, USA}
\author{A.~A.~Stark}
\affiliation{Center for Astrophysics $\vert$ Harvard \& Smithsonian, 60 Garden Street, Cambridge, MA 02138, USA}
\author{E.~Suchyta}
\affiliation{Computer Science and Mathematics Division, Oak Ridge National Laboratory, Oak Ridge, TN 37831}
\author{M.~E.~C.~Swanson}
\affiliation{National Center for Supercomputing Applications, 1205 West Clark St., Urbana, IL 61801, USA}
\author{G.~Tarle}
\affiliation{Department of Physics, University of Michigan, Ann Arbor, MI 48109, USA}
\author{D.~Thomas}
\affiliation{Institute of Cosmology and Gravitation, University of Portsmouth, Portsmouth, PO1 3FX, UK}
\author{C.~To}
\affiliation{Department of Physics, Stanford University, 382 Via Pueblo Mall, Stanford, CA 94305, USA}
\affiliation{Kavli Institute for Particle Astrophysics \& Cosmology, P. O. Box 2450, Stanford University, Stanford, CA 94305, USA}
\affiliation{SLAC National Accelerator Laboratory, Menlo Park, CA 94025, USA}
\author{D.~L.~Tucker}
\affiliation{Fermi National Accelerator Laboratory, P. O. Box 500, Batavia, IL 60510, USA}
\author{T.~N.~Varga}
\affiliation{Max Planck Institute for Extraterrestrial Physics, Giessenbachstrasse, 85748 Garching, Germany}
\affiliation{Universit\"ats-Sternwarte, Fakult\"at f\"ur Physik, Ludwig-Maximilians Universit\"at M\"unchen, Scheinerstr. 1, 81679 M\"unchen, Germany}
\author{R.~H.~Wechsler}
\affiliation{Department of Physics, Stanford University, 382 Via Pueblo Mall, Stanford, CA 94305, USA}
\affiliation{Kavli Institute for Particle Astrophysics \& Cosmology, P. O. Box 2450, Stanford University, Stanford, CA 94305, USA}
\affiliation{SLAC National Accelerator Laboratory, Menlo Park, CA 94025, USA}
\author{Z.~Zhang}
\affiliation{Kavli Institute for Cosmological Physics, University of Chicago, Chicago, IL 60637, USA}

\collaboration{DES \& SPT Collaborations}

\begin{abstract}
We perform a joint analysis of the counts of \redmapper\ clusters selected from the Dark Energy Survey (DES) Year 1 data and multi-wavelength follow-up data collected within the $2500 \deg^2$ South Pole Telescope (SPT) Sunyaev-Zel'dovich (SZ) survey. The SPT follow-up data, calibrating the richness--mass relation of the optically selected \redmapper\ catalog, enable the cosmological exploitation of the DES cluster abundance data. 
To explore possible systematics related to the modeling of projection effects, we consider two calibrations of the observational scatter on richness estimates: a simple Gaussian model which accounts only for the background contamination (BKG), and a model which further includes contamination and incompleteness due to projection effects (PRJ).
Assuming either a $\Lambda$CDM+$\sum m_\nu$ or $w$CDM+$\sum m_\nu$ cosmology, and for both scatter models, we derive cosmological constraints consistent with multiple cosmological probes of the low and high redshift Universe, and in particular with the SPT cluster abundance data. This result demonstrates that the DES Y1 and SPT cluster counts provide consistent cosmological constraints, if the same mass calibration data set is adopted. It thus supports the conclusion of the DES Y1 cluster cosmology analysis which interprets the tension observed with other cosmological probes in terms of systematics affecting the stacked weak lensing analysis of optically--selected low--richness clusters. Finally, we analyse the first combined optically-SZ selected cluster catalogue obtained by including the SPT sample above the maximum redshift probed by the DES Y1 \redmapper\ sample ($z=0.65$).
Besides providing a mild improvement of the cosmological constraints, this data combination serves as a stricter test of our scatter models: the PRJ model, providing scaling relations consistent between the two abundance and  multi-wavelength follow-up data, is favored over the BKG model.
\end{abstract}

\maketitle


\section{Introduction}
\label{sec:intro}
Tracing the highest peaks of the matter density field, galaxy clusters are a sensitive probe of the growth of structures \citep[see e.g.][for reviews]{Allen2011,Kravtsov2012}. 
In particular, the abundance of galaxy clusters as a function of mass and redshift has been used over the last two decades to place independent and competitive constraints on the density and amplitude of matter fluctuations, as well as dark energy and modified gravity models \citep[e.g.][]{Vikh2009,Rozo2010,Mantz2015,PlanckSZ2016,Bocquet2018,costanzietal18b,desy1cl}.
Thanks to the increasing number of wide area surveys at different wavelengths --- e.g. in the optical the Sloan Digital Sky Survey\footnote{https://www.sdss.org/} and the Dark Energy Survey\footnote{https://www.darkenergysurvey.org} (DES), in the microwave Planck\footnote{https://www.cosmos.esa.int/web/planck}, South Pole Telescope\footnote{https://pole.uchicago.edu/} (SPT) and Atacama Cosmology Telescope\footnote{https://act.princeton.edu/}, and in the X-ray eROSITA\footnote{https://www.mpe.mpg.de/eROSITA} ---
cluster catalogs have grown in size by an order of magnitude compared to early studies, extending to lower mass systems and/or to higher redshifts.
Despite this improved statistic, the constraining power of current cluster abundance studies is limited by the uncertainty in the calibration of the relation between cluster mass and the observable used as mass proxy \citep[see e.g.][]{Pratt2019}.
In general, the observable--mass relation (or OMR) can be calibrated either using high-quality X-ray, weak lensing and/or spectroscopic follow-up data for a representative sub-sample of clusters \citep[e.g.][]{Mantz2015,Bocquet2018,capasso2018}, or, if wide area imaging data are available, exploiting the noisier weak lensing signal measured for a large fraction of the detected clusters \citep[e.g.][]{costanzietal18b,murata2019,desy1cl}.
Depending on the methodology adopted the mass estimates can be affected by different sources of systematics: e.g. violation of the hydrostatic or dynamical equilibrium when relying on X-ray or spectroscopic follow-up data, respectively, or shear and photometric biases in weak lensing analyses. 
The calibration of the scaling relation is further hampered by the cluster selection and correlations between observables, which, if not properly modeled, can lead to large biases in the inferred parameters.
The recent analysis of the optical cluster catalog extracted from the DES year 1 data (Y1), which combines cluster abundance and stacked weak lensing data, exemplifies such limitations \citep[][hereafter DES20]{desy1cl}. 
The \citetalias{desy1cl} analysis results in cosmological posteriors in tension with multiple cosmological probes. The tension is driven by low richness systems,
and has been interpreted in terms of an unmodeled systematic affecting the stacked weak lensing signal of optically selected clusters.

A possible route to improve our control over systematics
relies on the combination of mass--proxies observed at different wavelengths, and thus not affected by the same sources of error.
Even more advisable would be the combination of cluster catalogs selected at different wavelengths which would enable the full exploitation of the cosmological content of current and future cluster surveys.
The DES and SPT data provides such an opportunity thanks to the large area shared between the two footprints and the high quality of the photometric and millimeter-wave data, respectively. 
Moreover, the X-ray and weak lensing follow-up data collected within the SPT survey
provide an alternative data set to the stacked weak lensing signal adopted in \citetalias{desy1cl} to constrain the observable--mass scaling relations, that has already been extensively vetted \citep{SPTSZ2016,Bocquet2018}.
The goal of this study is twofold: i) reanalyze the DES Y1 cluster abundance data adopting the SPT follow-up data to calibrate the observable--mass relation(s), and ii) provide a first case study for the joint analysis of cluster catalogs selected at different wavelengths.
In turn, this serves as independent test of the conclusions drawn in \citetalias{desy1cl}; secondly, combining the abundance data of the two surveys, we explore the possible cosmological gain given by the joint analysis of the two catalogs and exploit the complementary mass and redshift range probed by the two surveys to test the internal consistency of the data sets.
Concerning this last point, we consider two calibrations of the observational noise on richness estimates with the aim of assessing possible model systematics induced by a too simplistic modeling of the relation between richness and mass.

The paper is organized as follows: In section \ref{sec:datasets} we present the data sets employed in this work. Section \ref{sec:model} introduces the methodology used to analyze the data. We present our results and discuss their implication in section \ref{sec:res}. Finally we draw our conclusions in section \ref{sec:concl}.


\begin{figure}
 	\includegraphics[width= \linewidth]{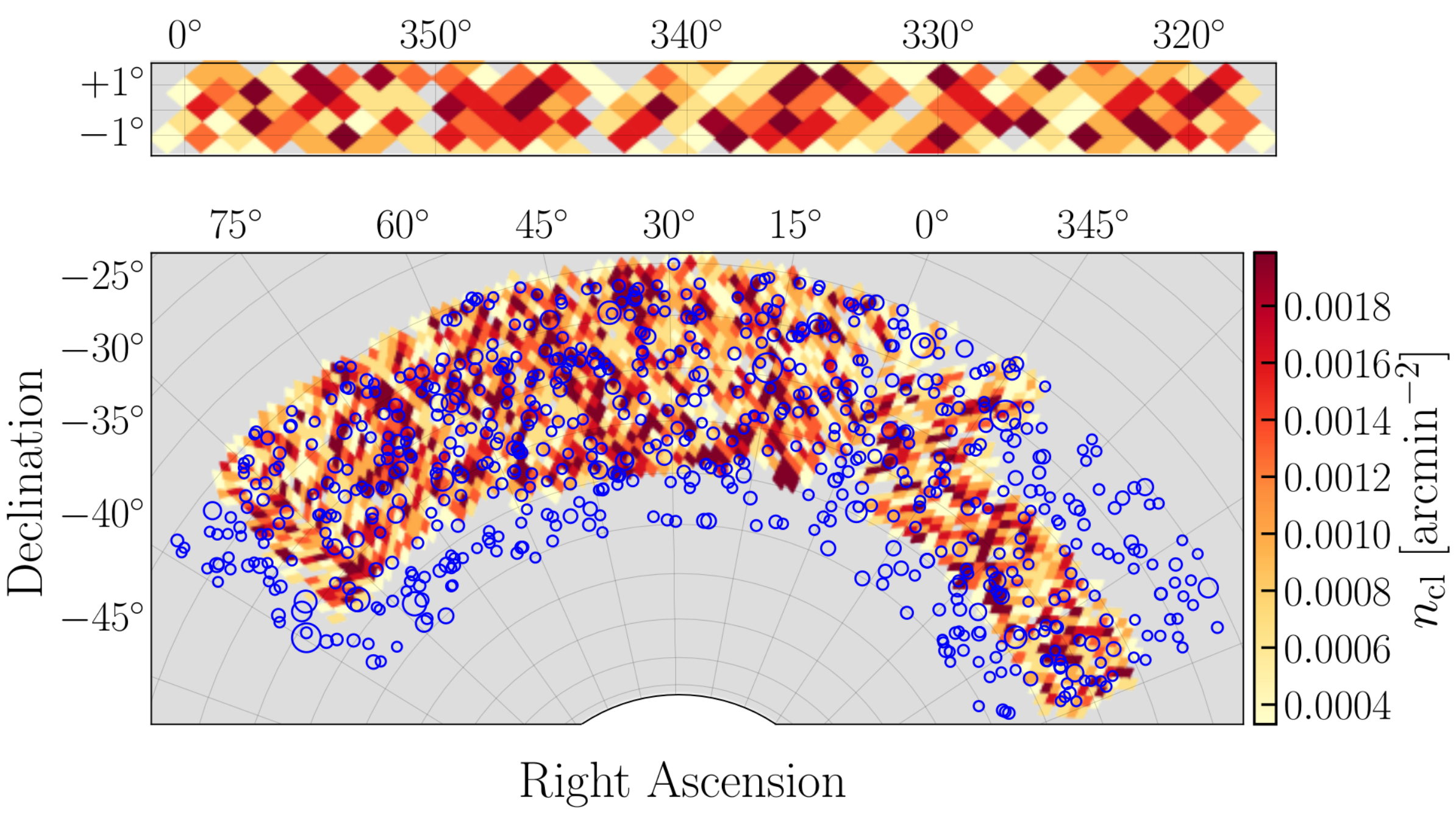}
	\caption{The DES Y1 \redmapper\ cluster density ($\lambda >20$) over the two non-contiguous regions of the Y1 footprint: the Stripe 82 region (116 deg$^2$; \textit{upper} panel) and the SPT region (1321 deg$^2$; \textit{lower} panel). In the \textit{lower} panel, we also show the locations of the SPT-SZ $2500 \deg^2$ clusters ($\xi>5$) in \textit{blue} circles with sizes proportional to the detection significance.
	}
	\label{fig:footprint}
\end{figure}

\section{Data}
\label{sec:datasets}
In this work we combine cluster abundance data from the DES Y1 \redmapper\ optical cluster catalog \citep[DES Y1 RM;][]{desy1cl}, with multi-wavelength data collected within the 2500 deg$^2$ SPT-SZ cluster survey \citep[SPT-SZ;][]{bleemetal15,Bocquet2018}.
Exploiting the large overlap ($\sim 1300 \deg^2$) of the DES Y1 and SPT-SZ survey footprints, we aim to use the SPT-SZ multi-wavelength data to calibrate the observable--mass relation of \redmapper\ clusters, which in turn enables the derivation of cosmological constraints from the DES Y1 abundance data.
Below we present a summary of the data sets employed in this work. To build our data vectors we follow the prescriptions adopted in \citetalias{desy1cl} and \cite{Bocquet2018} (hereafter B19) and refer the reader to the original works for further details.

\subsection{DES Y1 \redmapper\ Cluster Catalog}
\label{sec:redmapper}
The DES Y1 \redmapper\ clusters are extracted from the DES Y1 photometric galaxy catalog \citep{Y1gold}. The latter is based on the photometric data collected by the DECam during the Year One (Y1) observational season (from August 31, 2013 to February 9, 2014) over $\sim$1800 ${\rm deg}^2$ of the southern sky in the $g$, $r$, $i$, $z$ and $Y$ bands. Galaxy clusters are selected through the \redmapper\ photometric cluster finding algorithm that identifies galaxy clusters as overdensities of red-sequence galaxies \citep{Rykoff2014,rykoffetal16}. 
\redmapper\ uses a matched filter approach to estimate the membership probability of each red-sequence galaxy brighter than a specified luminosity threshold, $\Lmin(z)$, within an empirically calibrated cluster radius ($R_\lambda = 1.0\ \hinv\ \Mpc (\lob/100)^{0.2}$). The sum of these membership probabilities is called richness, and is denoted as $\lob$. Along with the richness, \redmapper\ estimates the photometric redshift of the identified galaxy clusters. Typical DES Y1 cluster photometric redshift uncertainties are $\sigma_z/(1+z)\approx 0.006$ with negligible bias ($|\Delta z| \leq 0.003$). The photometric redshift errors are both redshift and richness dependent.
To determine candidate central galaxies the \redmapper\ algorithm iteratively self-trains a filter that relies on galaxy brightness, cluster richness, and local galaxy density. The algorithm centers the cluster on the most likely candidate central galaxy which is not necessarily the brightest cluster galaxy. \cite{zhangetal19} studied the centering efficiency of the \redmapper\ algorithm using X-ray imaging and found that the fraction of correctly centered clusters is $\fcen=0.75 \pm 0.08$ with no significant dependence on richness.
 
Following \citetalias{desy1cl}, we use for the cluster count analysis the DES Y1 \redmapper\ volume-limited catalog with $\lob \geq 20$, in the redshift interval $z\in[0.2,0.65]$
, with a total of 6504 clusters\footnote{The \redmapper\ catalog can be found here: \url{https://des.ncsa.illinois.edu/releases/y1a1/key-catalogs/key-redmapper}}.
Galaxy clusters are included in the volume-limited catalog if the cluster redshift $z \leq \zmax(\hatn)$, where $\zmax(\hatn)$ is the maximum redshift at which galaxies at the luminosity threshold $\Lmin(z)$ are still detectable in the DES Y1 at $10\sigma$. Figure~\ref{fig:footprint} shows the cluster density in the two non-contiguous regions of the DES Y1 \redmapper\ cluster survey considered in this work. The lower panel, dubbed the SPT region, corresponds to the $\sim 1300 \deg^2$ overlapping area between the SPT-SZ and DES Y1 survey footprints.

Accordingly with the binning scheme adopted in \citetalias{desy1cl}, we split our cluster sample in four richness bins and three redshift bins as listed in Table~\ref{tab:abundances}. Moreover, we correct the cluster count data for miscentering effects following the prescription of \citetalias{desy1cl}. Briefly, cluster miscentering tends to bias low the richness estimates and thus the abundance data, introducing covariance amongst neighboring richness bins. The correction and covariance matrix associated with this effect are estimated in \citetalias{desy1cl} through Monte Carlo realizations of the miscentering model of \cite{zhangetal19}. The corrections derived for each richness/redshift bin are of the order of $\approx 3\%$ with an uncertainty of $\approx 1.0\%$ (see Table~\ref{tab:abundances}).


\begin{table*}
  \caption{Number of galaxy clusters in each richness and redshift bin for the DES Y1 \redmapper\ catalog.
  Each entry takes the form $N (N) \pm \Delta N\ {\rm stat}\ \pm \Delta N\ {\rm sys}$. The first error bar is the statistical uncertainty in the number of galaxy clusters in that bin given by the sum of a Poisson and a sample variance term.  The number between parenthesis and the second error bar correspond to the number counts corrected for the miscentering bias factors and the corresponding uncertainty (see section \ref{sec:redmapper}).}
  \label{tab:abundances}
    \begin{tabular}{cccc}
		$\lob$ & $z\in[0.2,0.35)$ & $z\in[0.35,0.5)$ & $z\in[0.5,0.65)$  \\ \hline \vspace{-2mm} \\
		$[20,30)$ & 762 (785.1) $\pm$ 54.9 $\pm$ 8.2 & 1549 (1596.0) $\pm$ 68.2 $\pm$ 16.6 & 1612 (1660.9) $\pm$ 67.4 $\pm$ 17.3\\
		$[30,45)$ & 376 (388.3) $\pm$ 32.1 $\pm$ 4.5 & 672 (694.0) $\pm$ 38.2 $\pm$ 8.0 & 687 (709.5) $\pm$ 36.9 $\pm$ 8.1\\
		$[45,60)$ & 123 (127.2) $\pm$ 15.2 $\pm$ 1.6 & 187 (193.4) $\pm$ 17.8 $\pm$ 2.4 & 205 (212.0) $\pm$ 17.1 $\pm$ 2.7\\
		$[60,\infty)$ & 91 (93.9) $\pm$ 14.0 $\pm$ 1.3 & 148 (151.7) $\pm$ 15.7 $\pm$ 2.2 & 92 (94.9) $\pm$ 14.2 $\pm$ 1.4\\
    \end{tabular}
\end{table*}


\begin{table}
  \centering
  \caption{Summary of the SPT-SZ cluster data used in this analysis split in mass--calibration data (SPT-OMR), and abundance data (SPT-NC). For the SPT-OMR data we specify in the third column the number of clusters with a specific follow-up measurement (see section \ref{sec:sptsz} for details). Note that a cluster might have more than one follow-up measurement.}
  \label{tab:spt}
    \begin{tabular}{cccc}
		Data set & Number of Clusters &  Follow-up & $z$-cut \\ \hline \vspace{-2mm} \\
		& & WL: 32 & $z>0.25$ \\
		SPT-OMR & 187 & $\lambda$: 129 & $0.25<z<0.65$ \\
		& & X-ray: 89 & $z>0.25$ \vspace{1.0mm} \\
		
		SPT-NC & 141 & & $z>0.65$
    \end{tabular}
\end{table}


\begin{figure}
\begin{center}
    \includegraphics[width=0.45 \textwidth]{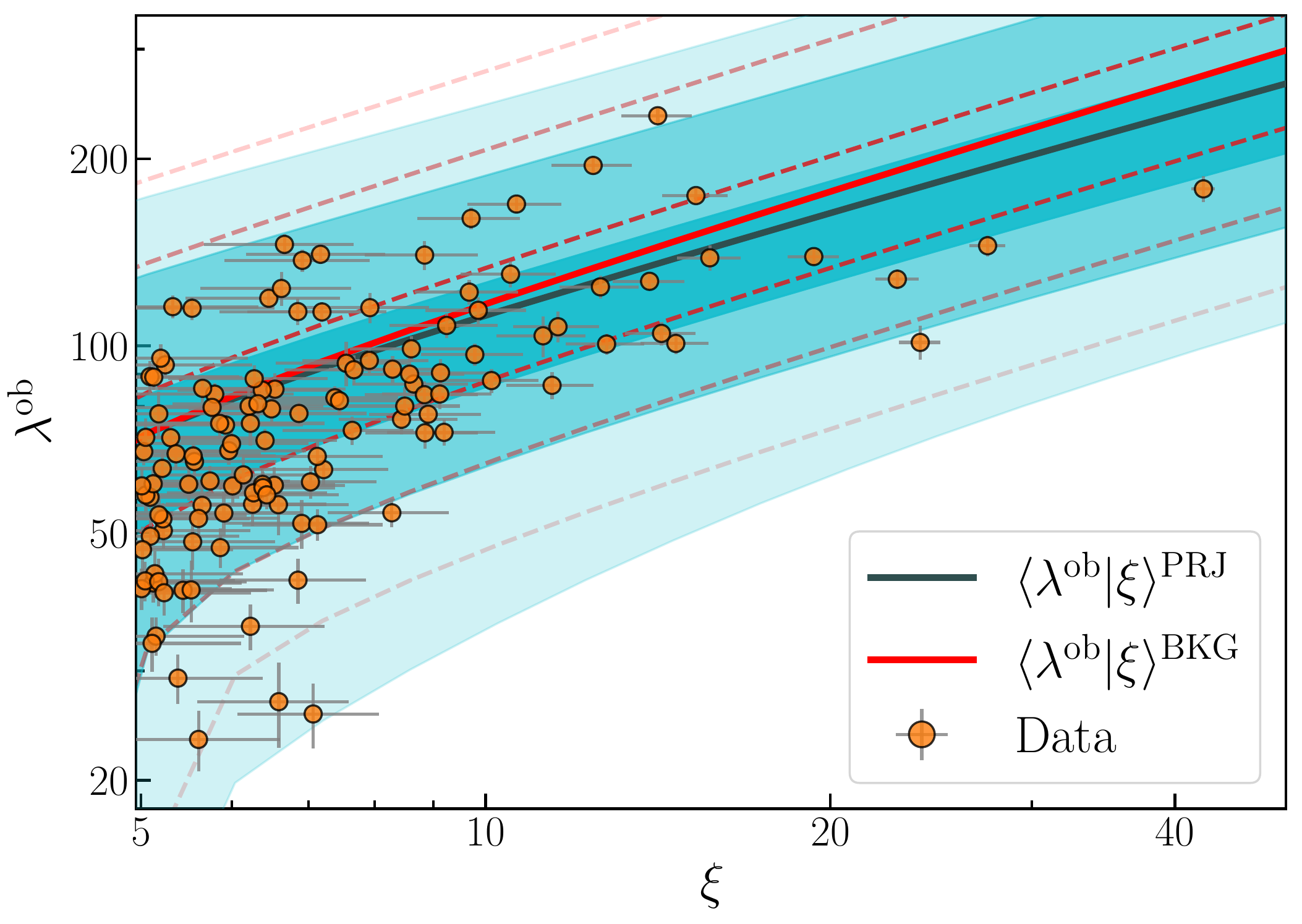}
\end{center}
\caption{Richness-SZ scaling relation for the DES Y1 RM-SPT SZ matched sample. The data points represent the observed values for the two mass proxies with the corresponding observational errors. The solid lines correspond to the mean relations derived from the DES-NC+SPT-OMR analysis adopting either the BKG (\textit{red}) or PRJ (\textit{dark cyan}) calibration for $P(\lob|\lambda)$ (see section \ref{sec:mor}). The dashed lines and bands represent, from the bottom to the top, the 0.13, 2.5, 16, 68, 97.5 and 99.87 percentile of the distributions for the BKG and PRJ models, respectively.}
\label{fig:lob_xi_rel}
\end{figure}


\subsection{SPT-SZ 2500 Cluster Catalog and Follow-Up Data}
\label{sec:sptsz}
Galaxy clusters are detected in the millimeter wavelength via the thermal Sunyaev-Zel’dovich signature \citep[SZ,][]{SZ1972} which arises from the inverse Compton scattering of CMB photons with hot electrons in the intracluster medium (ICM). The SPT-SZ survey observed the millimeter sky in the 95, 150, and 220 GHz bands over a contiguous 2500 deg$^2$ area reaching a fiducial depth of $ \leq18 \mu $K-arcmin in the 150 GHz band. Galaxy clusters are extracted from the SPT-SZ maps using a multi-scale matched-filter approach \citep{Melin2006} applied to the 95 and 150 GHz bands data as described in \cite{Williamson2011,Reichardt2013,bleemetal15}. For each cluster candidate, corresponding to a peak in the matched-filtered maps, the SZ observable $\xi$ is defined as the maximum detection significance over twelve equally spaced filter scales ranging from $0.'25$ to $3'$ \citep{bleemetal15}.
The SPT-SZ cosmological sample consists of 365 candidates with $\xi > 5$ and redshift $z>0.25$\footnote{Below z=0.25 the $\xi$-mass relation breaks due to confusion with the primary CMB fluctuations} (\textit{blue} circles in figure \ref{fig:footprint}). 
Of these: 343 clusters are optically confirmed and have redshift measurements, 89 have X-ray follow-up measurements with {\it Chandra} \citep{McDonald2013,McDonald2017}, 32 have weak lensing shear profile measurements from ground-based observations with Magellan/Megacam \citep[19 clusters;][]{Dietrich2019} and from space observations with the Hubble Space Telescope \citep[13 clusters;][]{Schrabback2018}.

Finally, to calibrate the \redmapper\ richness--mass relation we assign richnesses to the SPT-SZ clusters by cross-matching the two catalogs. To mitigate the impact of the optical selection we consider for the matching procedure all the clusters with $\lob \geq 5$ in the DES Y1 \redmapper\ volume-limited catalog. The match is performed following the criterion adopted in \cite{Bleem2020}; see also \cite{Saro2015} for an analogous study. Specifically: i) we sort the SPT-SZ and DES Y1 RM sample in descending order according to their selection observable, $\xi$ and $\lob$; ii) starting with the SPT-SZ cluster with the largest $\xi$, we match the system to the richest DES Y1 RM cluster within a projected radius of $1.5\, \Mpc$ and redshift interval $\delta_z=0.1$; iii) we remove the matched DES Y1 RM cluster from the list of possible counterparts and move to the next SPT-SZ system in the ranked list iterating step ii) until all the SPT-SZ clusters have been checked for a match.

We match all the 129 optically confirmed SPT-SZ clusters with $\xi > 5$ and $z>0.25$ that are in the proper redshift range and that lie in the DES Y1 footprint. The remaining 214 non-matched systems reside either in masked regions of the DES Y1 footprint or at redshifts larger than the local maximum redshift $\zmax(\hatn)$ of the DES Y1 RM volume-limited catalog.
Figure \ref{fig:lob_xi_rel} shows the $\lob$ distribution of the matched sample as a function of the SZ detection significance. The median of the distribution is $\lob=78$, while $68\%$ and $95\%$ of the matched sample resides above richness $\lob>60$ and $\lob>37$, respectively. 
To assess the probability of false association we repeat the matching procedure with $1000$ randomized DES Y1 RM catalogs and compute the fraction of times that an SPT-SZ system is associated with a random \redmapper\ cluster with $\lambda \geq \lob$. We find this probability to be less than $0.2\%$ for all the SPT-SZ matched systems, and thus we neglect it for the rest of the analysis.

We also explore the possible cosmological gain given by the inclusion of the number count data from the SPT-SZ catalog. When included, we only consider SPT-SZ clusters above redshift $0.65$ --- the redshift cut adopted for the DES Y1 \redmapper\ catalog --- corresponding to $40\%$ of the whole SPT-SZ sample. This redshift cut ensures the independence of DES Y1 RM and SPT-SZ abundance data, which allows a straightforward combination of the two data sets.      

A summary of the SPT-SZ data employed in this analysis can be found in Table \ref{tab:spt}.


\section{Analysis Method}
\label{sec:model}
Operatively, we can split our data set in three sub-samples and corresponding likelihoods: i) the DES Y1 RM abundance data (DES-NC), ii) the SPT-SZ multi-wavelength data (SPT-OMR) and iii) the SPT-SZ abundance data at $z>0.65$ (SPT-NC).
Our theoretical model for the DES Y1 RM number counts is the same as that described in detail in \cite{costanzietal18b} and \citetalias{desy1cl}, while for the analysis of the SPT-SZ abundance and multi-wavelength data we rely on the model presented in \citetalias{Bocquet2018}. Here we only provide a brief summary of these methods and refer the reader to the original works for further details. 
Throughout the paper, all quantities labeled with ``ob'' denote quantities inferred from observation, while $P(Y|X)$ denotes the conditional probability of $Y$ given $X$.  All masses are given in units of ${\rm M}_\odot / h$, where $h=H_0/100\vel \Mpc^{-1}$, and refer to an overdensity of 500 with respect to the critical density. We use "$\log$" and "$\ln$" to refer to the logarithm with base $10$ and $e$, respectively.


\subsection{Observable-Mass Relations Likelihood}
\label{sec:mor}
The SPT-SZ multi-wavelength data comprises four mass proxies: the SZ detection significance $\xi$, the richness $\lob$, the X-ray radial profile $Y_X\ob$, and the reduced tangential shear profile $g_t(\theta)$. The corresponding mean observable--mass relations for the intrinsic quantities -- $\zeta$, $\lambda$, $Y_X$, $\mwl$ --  are parameterized as follows:
\begin{multline}
    \label{eqn:zeta_m}
    \langle \ln \zeta \rangle = \ln (\gamma_f A_\sz) + B_\sz \ln \left( \frac{M}{3 \times 10^{14} \msunh} \right) + \\ + C_\sz \ln \left( \frac{E(z)}{E(0.6)}\right)
\end{multline}
\begin{multline}
    \label{eqn:lam_m}
    \langle \ln \lambda \rangle= \ln (A_\lambda) + B_\lambda \ln \left( \frac{M}{3 \times 10^{14} \msunh} \right) + \\ + C_\lambda \ln \left( \frac{1+z}{1+0.45}\right)
\end{multline}
\begin{multline}
    \label{eqn:Yx_m}
    \ln \left( \frac{M}{5.86 \times 10^{13} \msunh} \right)= \ln (A_{Y_X}) + B_{Y_X} \langle \ln Y_X \rangle + \\ + B_{Y_X}  \ln \left( \frac{(h/0.7)^{5/2}}{3 \times 10^{14} \msun {\rm keV}} \right) + C_{Y_X} \ln E(z)
\end{multline}
\begin{equation}
    \label{eqn:mwl_m}
    \langle \ln \mwl \rangle= \ln b_{\rm WL} + \ln M \, ,
\end{equation}
where $\gamma_f$ in equation \ref{eqn:zeta_m} depends on the position of the SPT-SZ cluster and accounts for the variation of survey depth over the SPT footprint \citep{SPTSZ2016}, while $E(z)=H(z)/H_0$. For each scaling relation we fit for the amplitude, slope, and redshift evolution (see Table \ref{tab:parameters}), but for the weak lensing mass, $\mwl$, which we assume to be simply proportional to the true halo mass accordingly to the simulation-based calibration of \citetalias{Bocquet2018}.

We assume the logarithm of our four intrinsic observables, $\ln  \mathcal{O}$, to follow a multivariate Gaussian distribution with intrinsic scatter parameters $D_\mathcal{O}$, and correlation coefficients $\rho( \mathcal{O}_i;\mathcal{O}_j)$:
\begin{equation}
    \label{eqn:Plnob_M}
    P(\ln  \mathcal{O}|M,z) =   \mathcal{N} \left( \langle \ln  \mathcal{O} \rangle, {\bm C} \right) \, ,
\end{equation}
where the covariance matrix elements read $C_{ij}=\rho( \mathcal{O}_i;\mathcal{O}_j) D_\mathcal{O}^i D_\mathcal{O}^j$ and $\rho( \mathcal{O}_i;\mathcal{O}_i)=1$.
All the intrinsic scatters are described by a single parameter $D_\mathcal{O}$ independent of mass and redshift, but the scatter on $\ln \lambda$ which includes a Poisson--like term --  $\sigma_{\ln  \lambda}^2=D^2_\lambda+(\langle \lambda(M) \rangle-1)/\langle \lambda(M) \rangle^2$ -- which does not correlate with the other scatter parameters. Finally, we set to zero the correlation coefficients between the $D_{Y_X}$ and the other scatter parameters. This approximation is justified by the fact that while the richness, SZ and weak lensing signal are sensitive to the projected density field along the line of sight of the system, the X-ray emission is mainly contributed by the inner region of the cluster. This approximation is also supported by the analysis of \citetalias{Bocquet2018} which obtained unconstrained posteriors peaked around zero for the X-ray correlation coefficients.
We explicitly verified that this approximation does not affect our results, while reducing noticeably the computational cost of the analysis.

To account for the observational uncertainties and/or biases, we consider the following conditional probabilities between the intrinsic cluster proxies and the actual observed quantities. For $\xi$, $Y_x$ and $\gamma_t(\theta)$ we follow the prescriptions outlined in \citetalias{Bocquet2018}, namely:
\begin{eqnarray}
    \label{eqn:Pxi_zeta}
    P(\xi|\zeta) &=&  \mathcal{N} \left(\sqrt{\zeta^2+3},1 \right) \\
    \label{eqn:PYob_Ytr}
    P(Y_X\ob|Y_X)&=&  \mathcal{N} \left(Y_X,\sigma_{Y_X}\ob \right) \, ,
\end{eqnarray}
where $\sigma_{Y_X}\ob$ is the uncertainty associated with the X-ray measurements \citep[see section 3.2.2 in][for further details]{Bocquet2018}.
The reduced tangential shear $g_t(\theta)$ is analytically related to the underlying halo mass $\mwl$ assuming a Navarro-Frenk-White (NFW) halo profile \citep{NFW}, a concentration--mass relation, and using the observed redshift distribution of source galaxies.
Deviation from the NFW profile, large-scale structure along the line of sight, miscentering and uncertainties in the concentration–mass relation, introduce bias and/or scatter on the estimated weak lensing mass, $\mwl$. As introduced in equation \ref{eqn:mwl_m}, we assume $\mwl$ to be proportional to the true halo mass, and use the simulation-based calibration of $b_{\rm WL}$ from \citetalias{Bocquet2018} to account for such effects (see their Section 3.1.2 and Table 1 for further details). In total the weak lensing (WL) modeling introduces six free parameters which account for the uncertainties in the determination of the systematics associated to the mean bias ($\delta_{\rm{WL,bias}}$, $\delta_{\rm{HST/MegaCam,bias}}$) and scatter ($\delta_{\rm{WL,scatter}}$, $\delta_{\rm{HST/MegaCam,scatter}}$) of the WL--mass scaling relation. Of these, two parameters are shared among the entire WL sample ($\delta_{\rm{WL,bias}}$, $\delta_{\rm{WL,scatter}}$), while the other two pairs are associated with the sub-sample observed with HST ($\delta_{\rm{HST,bias}}$, $\delta_{\rm{HST,scatter}}$) or Megacam ($\delta_{\rm{MegaCam,bias}}$, $\delta_{\rm{MegaCam,scatter}}$).

As for the uncertainty on the richness, many studies already highlight the importance of projection effects on richness estimates \citep[e.g.][]{farahietal16,zuetal17,buschwhite17,Murata2017,Wojtak2018,costanzietal18}. In this context, projection effects denote the contamination from correlated and uncorrelated structures along the line of sight due to the limited resolution that a photometric cluster finding algorithm can achieve along the radial direction. In this study we consider two prescriptions based on the model presented in \cite{costanzietal18}:
\begin{enumerate}
    \item{$P_{\rm bkg}(\lob|\lambda ) =  \mathcal{N} (\lambda,\sigma_{\lambda}^{\rm bkg} )$, which accounts only for the "background subtraction" scatter , $\sigma_{\lambda}^{\rm bkg}$, due to the misclassification of background galaxies as member galaxies and \textit{vice versa}, labelled BKG throughout the paper.}
    \item{$P_{\rm prj}(\lob|\lambda )$, defined in equation 15 of \cite{costanzietal18}, which includes, besides the "background subtraction" noise, the scatter due to projection and masking effects (PRJ, hereafter).}
\end{enumerate}
The approximated BKG model is derived from $P_{\rm prj}(\lob|\lambda )$ by setting to zero the fraction of clusters affected by projection and masking effects and corresponds to the model often adopted in literature \citep[e.g.][]{Saro2015,Murata2017,Bleem2020}.
PRJ is the model adopted in \citetalias{desy1cl}, and it has been calibrated by combining real data and simulated catalogs analysis. While being a more complete model which includes known systematics effects, its calibration, in part based on simulated catalogs, might be subject to biases.  
Comparisons of the results obtained with these two models are used to assess the capability of our simplest model (BKG) to absorb the impact of projection effects and, in turn, possible biases due to their incorrect calibration.

Putting all the above pieces together, the "observable--mass relation" likelihood for the SPT-SZ multi-wavelength data is given by:
\begin{equation}
\label{eqn:likeMOR}
    \ln \mathcal{L}^{\rm OMR}(\mathcal{O}\ob|{\bm \theta}) =  \sum_i \ln P(\lob_i,{Y_X}\ob_i,{g_\mathrm{t}}_i|\xi_i,z_i,{\bm \theta}) \, ,
\end{equation}
where ${\bm \theta}$ denotes the model parameters and the sum runs over all the SPT-SZ clusters with at least a follow-up measurement (besides $\xi$). Each term of the summation is computed as:
\begin{eqnarray}
   \label{eqn:Pob_xi}
   P(\lob,Y_X\ob,g_t|\xi,z,{\bm \theta}) \propto \int \de M \, \de \zeta \, \de \lambda \, \de Y_X \, \de \mwl & & \nonumber \\
   P(\xi|\zeta) P(\lob|\lambda)  P(Y_X\ob|Y_X) P(g_t|\mwl) & & \nonumber \\
   P(\zeta,\lambda,Y_X,\mwl|M,z) n(M,z) & & . 
\end{eqnarray}
In the above expression $n(M,z)$ represents the halo mass function for which we adopt the \cite{Tinker2008} fitting formula.
Following the original analyses of \citetalias{desy1cl} and \citetalias{Bocquet2018} we neglect the uncertainty on the halo mass function due to baryonic feedback effects, being the latter subdominant to the uncertainty on the cluster counts due to the mass calibration.
The proportionality constant is set by the normalization condition: $\int_5^\infty \de \lob \int \de \xi \de g_t \de Y_X\ob P(\lob,Y_X\ob,g_t|\xi,z)=1 $, where the lower limit is set by the $\lob \geq 5$ cut applied to the DES Y1 RM sample to match the catalogs. Finally, note that in the above expression only the integrals over the mass proxies for which we have a measurement need to be computed in practice. If no follow-up measurements are available for a SPT system the conditional probability reduces to one and thus can be omitted from the sum in equation \ref{eqn:likeMOR}. 


\subsection{Cluster Abundance Likelihoods}
\label{sec:nc}
The expected number of clusters observed with $\mathcal{O}\ob$ at redshift $z$, over a survey area $\Omega(z)$, is given by:
\begin{multline}
\label{eqn:nc}
    \langle N(\mathcal{O}\ob,z) \rangle = \int \de M n(M,z) {\Omega(z)} \left . \frac{\de V}{\de z \de \Omega} \right \vert_{z} \cdot \\
 \cdot \int \de \mathcal{O} P(\mathcal{O}\ob |\mathcal{O})P(\mathcal{O} |M,z) \, ,
\end{multline}
where  $\de V /(\de z \de \Omega)$ is the comoving volume element per unit redshift and solid angle, whereas the conditional probabilities for the observed and intrinsic mass proxies are those described in the previous section. 

The DES Y1 RM cluster abundance data are analyzed following the methodology adopted in \citetalias{desy1cl} where the number counts likelihood takes the form:
\begin{equation}
\label{eqn:likeNC_DES}
 \mathcal{L}^{\rm NC}_{\rm DES}({\bm N}_\Delta |{\bm \theta}  ) = \frac{ \exp \left[ -\frac{1}{2} \left( {\bm N}_\Delta -   \langle {\bm N}_\Delta \rangle \right)^T {\bm C}^{-1} \left( {\bm N}_\Delta -   \langle {\bm N}_\Delta \rangle \right) \right]}{\sqrt{(2 \pi)^{12} {\rm det}(\bm C)}} \, ,
\end{equation}
where ${\bm N}_\Delta$ and $ \langle {\bm N}_\Delta \rangle$ are respectively the abundance data (see Table \ref{tab:abundances}), and the expected number counts in bins of richness and redshift obtained by integrating equation \ref{eqn:nc} over the relevant $\lob$ and $z$ intervals. 
The covariance matrix ${\bm C}$ is modeled as the sum of three distinct contributions: i) the Poisson noise, ii) a sample variance term due to density fluctuations within the survey area and iii) a miscentering component (see section \ref{sec:redmapper}). The Poisson and sample variance contributions are computed analytically at each step of the chain following the prescription outlined in Appendix A of \cite{costanzietal18b}. At high richness, the Poisson term dominates the uncertainty, with sample variance becoming increasingly important at low richness \citep{hukravtsov03}.
Note that the large occupancy of all our bins --- our least populated bin contains 91 galaxy clusters ---justify the Gaussian approximation adopted for the Poisson component.

Following \citetalias{Bocquet2018}, we assume a purely Poisson likelihood for the SPT-SZ abundance data \citep{Cash1979}:
\begin{equation}
\label{eqn:likeNC_SPT}
 \ln \mathcal{L}^{\rm NC}_{\rm SPT}({\bm N} |{\bm \theta}  ) = \sum_i \ln \langle N(\xi_i,z_i) \rangle - \int_{0.65} \de z \int_5 \de \xi \langle N(\xi,z) \rangle \, ,
\end{equation}
where the sum runs over all the SPT-SZ clusters above the redshift and SZ significance cuts ($z_{\rm cut}=0.65$, $\xi_{\rm cut}=5$). Note that here we can safely neglect the sample variance contribution given
large cluster masses ($M \gtrsim 3 \times 10^{14} \msunh$) probed by the SPT-SZ survey (see \cite{hukravtsov03, Fumagalli}).


\subsection{Parameters Priors and Likelihood Sampling}
\label{sec:prior}

\begin{table}
    \centering
    \footnotesize
    \caption{Cosmological and model parameter posteriors: a range indicates a top-hat prior, while $\mathcal{N}(\mu,\sigma)$ stands for a Gaussian prior with mean $\mu$ and variance $\sigma^2$.}
    \label{tab:parameters}
   \begin{tabular}{lcc}
   
	Parameter			&	Description	& Prior	\\
    \hline \vspace{-2.0mm} \\
	$\Omega_m$		& Mean matter density 						& $[0.1,0.9]$ 		\vspace{0.5mm} \\
    $A_s$		& {\parbox{3.5cm}{Amplitude of the primordial curvature perturbations}}	& $[10^{-10},10^{-8}]$ 	\vspace{0.5mm} \\
    $h$ & Hubble rate & $[0.55,0.9]$    \vspace{0.5mm}  \\
    $\Omega_b h^2$ & Baryon density & $[0.020,0.024]$ \vspace{0.5mm}  \\
    $\Omega_\nu h^2$ &  Massive neutrinos energy density & $[0.0006,0.01]$   \vspace{0.5mm} \\
    $n_s$ & Spectral index & $[0.94,1.0]$ \vspace{0.5mm}  \\
    $w$ & Dark energy equation of state & $[-2.5,-0.33]$ \vspace{0.5mm}  \\
    \hline \vspace{-3mm}\\
    \multicolumn{3}{l}{SZ scaling relation}\\
    $A_\sz$	& Amplitude & $[1,10]$ 	\vspace{0.5mm} \\
    $B_\sz$		& Power-law index mass dependence  &$[1,2.5]$ 	\vspace{0.5mm} \\
    $C_\sz$		& Power-law index redshift evolution &$ [-1,2] $\vspace{0.5mm} \\
    $D_\sz$	& Intrinsic scatter			& $[0.01,0.5]$	\vspace{0.5mm} \\
    \hline \vspace{-3mm}\\
    \multicolumn{3}{l}{Richness scaling relation}\\
    $A_\lambda$	& Amplitude & $[20,120]$ \vspace{0.5mm} \\
    $B_\lambda$		& Power-law index mass dependence &$[0.4,2.0]$ 	\vspace{0.5mm} \\
    $C_\lambda$		& Power-law index redshift evolution &$ [-1,2] $ \vspace{0.5mm} \\
    $D_\lambda$	& Intrinsic scatter			& $[0.01,0.7]$ 		\vspace{0.5mm} \\   
    \hline \vspace{-3mm}\\
    \multicolumn{3}{l}{X-ray $Y_X$ scaling relation}\\
    $A_{Y_X}$	& Amplitude & $[1,10]$ 	\vspace{0.5mm} \\
    $B_{Y_X}$		& Power-law index mass dependence &$[1,2.5]$ 	\vspace{0.5mm} \\
    $C_{Y_X}$		& Power-law index redshift evolution &$ [-1,2] $\vspace{0.5mm} \\
    $D_{Y_X}$	& Intrinsic scatter			& $[0.01,0.5]$		\vspace{0.5mm} \\    
    $\de\ln Y_X/\de\ln r$	& Radial slope $Y_X$ profile & $\mathcal{N}(1.12,0.23)$	\vspace{0.5mm} \\    
    \hline \vspace{-3mm}\\
    \multicolumn{3}{l}{$\mwl$ scaling relation}\\
    $\delta_{\rm{WL,bias}}$	& Uncertainty on WL bias  & $\mathcal{N}(0,1)$ 	\vspace{0.5mm} \\
    $\delta_{\rm{bias}}^{\rm HST/MegaCam}$	&  {\parbox{3.5cm}{HST/MegaCam uncertainty on WL bias}}  & $\mathcal{N}(0,1)$ 	\vspace{0.5mm} \\
    $\delta_{\rm{WL,scatter}}$		& Uncertainty on intrinsic scatter &$\mathcal{N}(0,1)$		\vspace{0.5mm} \\
    $\delta_{\rm{scatter}}^{\rm HST/MegaCam}$	& {\parbox{3.5cm}{HST/MegaCam uncertainty on scatter due to uncorrelated LSS}} &$\mathcal{N}(0,1)$	\vspace{4.0mm} \\
    \hline \vspace{-3mm}\\   
    \multicolumn{3}{l}{Correlation coefficients between scatters}\\
    $\rho({\rm SZ,WL})$	& Correlation coefficient SZ-WL & $[-1,1]$ 	\vspace{0.5mm} \\
    $\rho({\rm SZ,}\lambda)$		& Correlation coefficient SZ-$\lambda$ & $[-1,1]$ 	\vspace{0.5mm} \\
    $\rho({\rm WL,}\lambda)$		& Correlation coefficient WL-$\lambda$ & $[-1,1]$ 	\vspace{0.5mm} \\
     & Determinant OMR matrix (eq.~\ref{eqn:Plnob_M}) & ${\rm det}|{\bm C}|>0$ \vspace{0.5mm} \\
    \hline \vspace{-3mm}\\
    \end{tabular}
\end{table}

The cosmological and model parameters considered in this analysis are listed in Table~\ref{tab:parameters} along with their priors.
Our reference cosmological model is a flat $\Lambda$CDM model with three degenerate species of massive neutrinos ($\Lambda$CDM+$\sum m_\nu$), for a total of six cosmological parameters: $\Omega_m$, $A_s$, $h$, $\Omega_b h^2$, $\Omega_\nu h^2$, $n_s$. Being that our data set is insensitive to the optical depth to reionization, we fix $\tau=0.078$. We also consider a $w$CDM+$\sum m_\nu$ model where the dark energy equation of state parameter $w$ is let free to vary in the range $[-2.5,-0.33]$.
The four observable--mass scaling relations considered in this work comprise 19 model parameters. Besides those already introduced in section \ref{sec:mor}, the $Y_X$ scaling relation has the additional parameter $(\de\ln Y_X/\de\ln r)$ -- the measured radial slope of the $Y_X$ profile -- which allows to re-scale and compare the measured and predicted $Y_X$ profiles at a fixed fiducial radius  \citep[see section 3.2.2 of ][for additional details]{Bocquet2018}. 
The parameters ranges and priors match those used in \citetalias{Bocquet2018}, apart from the richness--mass scaling relation parameters, which were not included in the \citetalias{Bocquet2018} analysis, and for which we adopt flat uninformative priors.
The parameter ranges for $\Omega_b h^2$ and $n_s$ are chosen to roughly match the $5\sigma$ credibility interval of the Planck constraints \citep{Planck2018}, while the lower limit adopted for $\Omega_\nu h^2$ corresponds to the minimal total neutrino mass allowed by oscillation experiments, 0.056 eV \citep{Tanabashi2018}.

We consider two different data combinations in this work. Our baseline data set is given by the combination of DES Y1 RM counts data and the SPT-SZ multi-wavelength data (DES-NC+SPT-OMR). Moreover, we explore the cosmological gain given by the further inclusion of the SPT-SZ abundance data (DES-NC+SPT-[OMR,NC]). The total log-likelihood is thus given by the sum of log-likelihoods corresponding to the data considered in each analysis. We remind here that the independence of the two abundance likelihoods is guaranteed by the redshift cut $z>0.65$ adopted for the SPT-SZ number count data which ensures the absence of overlap between the volume probed by the two abundance data sets.
The parameter posteriors are estimated within the {\tt cosmoSIS} package \citep{cosmosis} using the importance nested sampler algorithm {\tt MultiNest} \citep{multinest} with target error on evidence equal to 0.1 as convergence criterion.
. The matter power spectrum is computed at each step of the chain using the Boltzmann solver {\tt CAMB} \citep{camb}. To keep the universality of the Tinker fitting formula in cosmologies with massive neutrinos we adopt the prescription of \cite{Costanzi2013} neglecting the neutrino density component in the relation between scale and mass --- i.e. $M \propto (\rho_{\rm cdm} + \rho_{\rm b})R^3$ --- and using only the cold dark matter and baryon power spectrum components to compute the variance of the density field at a given scale, $\sigma^2(R)$.

%
\begin{figure*}
\begin{center}
    \includegraphics[width=\textwidth]{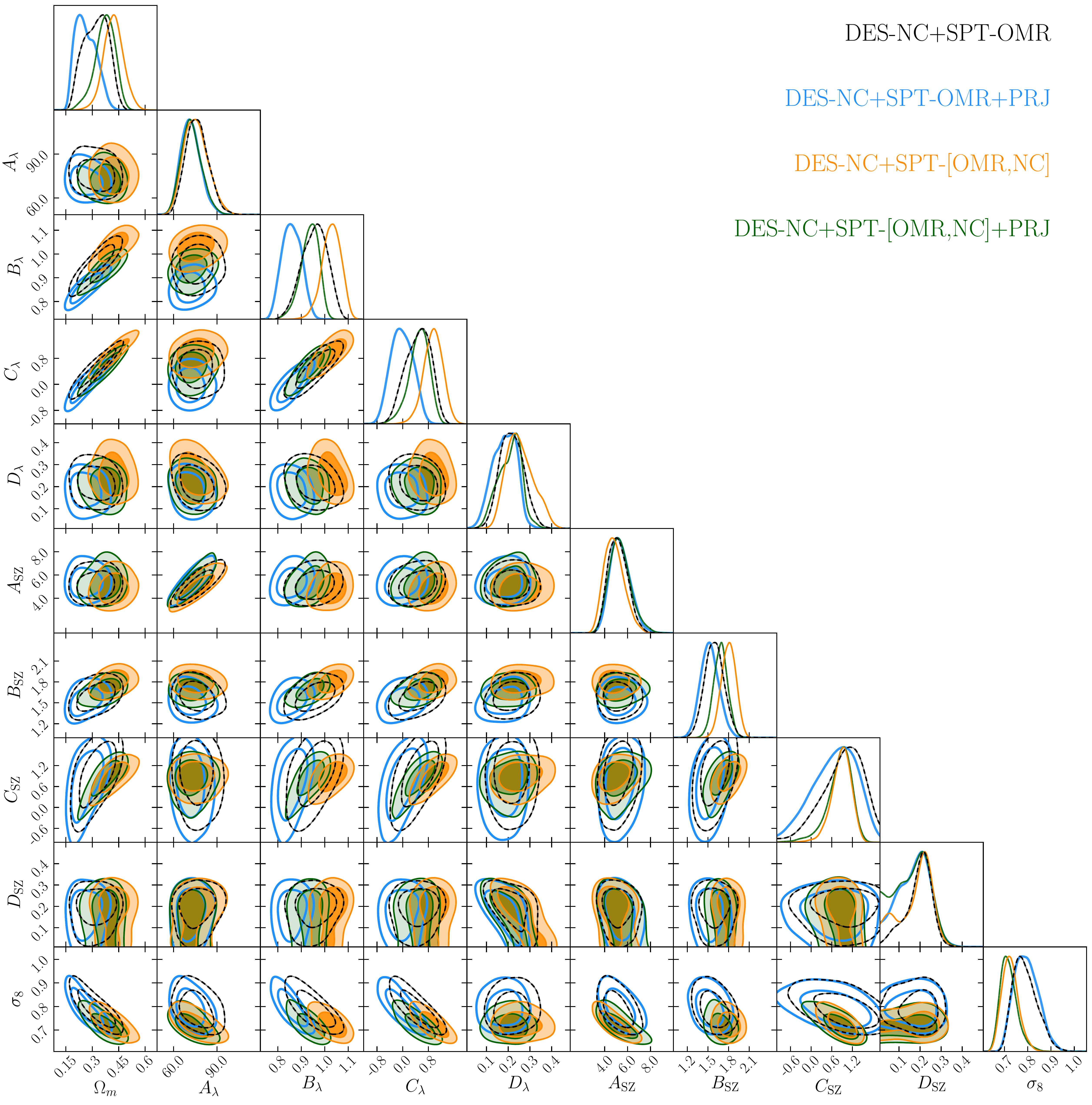}
\end{center}
\caption{Marginalized posterior distributions of the fitted parameters. The $2$D contours correspond to the $68\%$ and $95\%$ confidence levels of the marginalized posterior distribution. The description of the model parameters along with their posteriors are listed in Table \ref{tab:results}. Only parameters that are not prior dominated are shown in the plot.}
\label{fig:LCDM_constraints}
\end{figure*}
%

\begin{table*}
    \centering
    \footnotesize
    \caption{Cosmological and model parameter constraints obtained for the different models and data combinations considered in this work. For all the parameters we report the mean of the $1$-d marginalized posterior along with the $1$-$\sigma$ errors. We omit from this table parameters whose posteriors are equal to or strongly dominated by their priors. DES-NC, SPT-OMR and SPT NC stand for the different data set considered in the analyses, respectively: cluster counts from DES Y1 RM, multi-wavelength data from SPT-SZ, and abundance from the SPT-SZ cluster catalog above $z>0.65$. BKG and PRJ refer to the model adopted to describe the observational noise on the richness estimate (see section~\ref{sec:mor}).}
    
    \label{tab:results}
    \begin{tabular}{lcccccc} 

		&	\multicolumn{4}{c}{\LCDM+$\sum m_\nu$}& \multicolumn{2}{c}{$w$CDM+$\sum m_\nu$} \vspace{0.5mm} \\ \hline \vspace{-2.0mm} 	\\
     Data   & \multicolumn{2}{c}{DES-NC+SPT-OMR} & \multicolumn{2}{c}{DES-NC+SPT-[OMR,NC]} & DES-NC+SPT-OMR & DES-NC+SPT-[OMR,NC] \vspace{0.5mm} \\
    $P(\lob|\ltrue)$ model    & BKG & PRJ & BKG & PRJ & BKG & BKG \vspace{0.5mm} \\ \hline \vspace{-2.0mm} \\
	$\Omega_m$		& $0.322^{+0.079}_{-0.067}$	& $0.264^{+0.047}_{-0.073}$	& $0.420\pm 0.057$	& $0.372^{+0.064}_{-0.046}$	& $0.308^{+0.041}_{-0.054}$ & $0.362^{+0.044}_{-0.060}$	\vspace{0.5mm} \\
    $10^9 A_s $	& $2.38^{+0.42}_{-0.13}$	& $4.25^{+0.82}_{-0.20}$	& $1.21^{+0.21}_{-0.5}$	& $2.18^{+0.36}_{-0.92}$	& $1.64^{+0.25}_{-0.82}$ & $1.05^{+0.13}_{-0.40}$	\vspace{0.5mm} \\
    $h$ & $0.715^{+0.075}_{-0.091}$	& $0.677^{+0.045}_{-0.11}$	& $0.720\pm 0.075$	& $0.644^{+0.038}_{-0.076}$	& $0.765^{+0.12}_{-0.048}$ & $0.776^{+0.11}_{-0.046}$	\vspace{0.5mm} \\
    $\sigma_8$	& $0.790^{+0.038}_{-0.063}$	& $0.795^{+0.045}_{-0.059}$	& $0.725^{+0.030}_{-0.040}$	& $0.719^{+0.027}_{-0.042}$	& $0.808\pm 0.041$ & $0.771\pm 0.040$	\vspace{0.5mm} \\
    $S_8$	& $0.808^{+0.062}_{-0.049}$	& $0.736\pm 0.049$	& $0.854\pm 0.043$	& $0.796^{+0.048}_{-0.038}$	& $0.813^{+0.049}_{-0.044}$ & $0.842\pm 0.044$	\vspace{0.5mm} \\
    $w$	& $-1$	& $-1$	& $-1$	& $-1$	& $-1.76^{+0.46}_{-0.33}$ & $-1.95^{+0.48}_{-0.19}$	\vspace{0.5mm} \\
    $A_\sz$	& $5.18^{+0.74}_{-0.95}$	& $5.36^{+0.75}_{-1.0}$	& $4.84^{+0.72}_{-1.0}$	& $5.34^{+0.79}_{-1.0}$	& $4.16^{+0.60}_{-0.97}$ & $3.93^{+0.63}_{-0.91}$	\vspace{0.5mm} \\
    $B_\sz$	& $1.59\pm 0.14$	& $1.53^{+0.12}_{-0.14}$	& $1.80\pm 0.11$	& $1.69\pm 0.10$	& $1.67\pm 0.14$ & $1.85\pm 0.11$	\vspace{0.5mm} \\
    $C_\sz$	& $0.91^{+0.74}_{-0.42}$	& $0.68^{+0.78}_{-0.52}$	& $0.87^{+0.32}_{-0.24}$	& $0.82^{+0.41}_{-0.24}$	& $1.05^{+0.62}_{-0.42}$ & $1.33^{+0.26}_{-0.22}$	\vspace{0.5mm} \\
    $D_\sz$	& $0.193^{+0.074}_{-0.040}$	& $0.172^{+0.085}_{-0.070}$	& $0.182^{+0.098}_{-0.13}$	& $0.163^{+0.098}_{-0.074}$	& $0.193^{+0.082}_{-0.043}$ & $0.17^{+0.10}_{-0.14}$	\vspace{0.5mm} \\
    $A_\lambda$	& $76.3^{+6.9}_{-8.6}$	& $72.0^{+5.8}_{-7.7}$	& $75.6^{+7.0}_{-9.5}$	& $72.4^{+6.1}_{-7.9}$	& $66.1^{+6.1}_{-9.7}$ & $64.4^{+6.7}_{-9.2}$	\vspace{0.5mm} \\
    $B_\lambda$	& $0.957^{+0.059}_{-0.051}$	& $0.859\pm 0.040$	& $1.028^{+0.043}_{-0.037}$	& $0.935^{+0.045}_{-0.031}$	& $1.015^{+0.048}_{-0.037}$ & $1.058\pm 0.037$	\vspace{0.5mm} \\
    $C_\lambda$	& $0.48^{+0.45}_{-0.35}$	& $-0.02\pm 0.34$	& $0.95\pm 0.30$	& $0.51^{+0.35}_{-0.25}$	& $0.67\pm 0.34$ & $1.07\pm 0.30$	\vspace{0.5mm} \\
    $D_\lambda$	& $0.217^{+0.051}_{-0.058}$	& $0.183^{+0.064}_{-0.048}$	& $0.254^{+0.050}_{-0.075}$	& $0.207^{+0.061}_{-0.045}$	& $0.219\pm 0.058$ & $0.265^{+0.058}_{-0.082}$	\vspace{0.5mm} \\
    $A_{Y_X}$	& $6.91\pm 0.88$	& $6.41^{+0.76}_{-0.91}$	& $7.22\pm 0.72$	& $6.82\pm 0.72$	& $6.42^{+0.65}_{-0.84}$ & $6.87^{+0.67}_{-0.75}$	\vspace{0.5mm} \\
    $B_{Y_X}$	& $0.499^{+0.036}_{-0.049}$	& $0.519^{+0.040}_{-0.047}$	& $0.452^{+0.027}_{-0.036}$	& $0.479^{+0.030}_{-0.038}$	& $0.485^{+0.036}_{-0.046}$ & $0.446^{+0.028}_{-0.036}$	\vspace{0.5mm} \\
    $C_{Y_X}$	& $-0.47^{+0.20}_{-0.31}$	& $-0.43^{+0.24}_{-0.34}$	& $-0.35^{+0.11}_{-0.14}$	& $-0.37^{+0.12}_{-0.18}$	& $-0.52^{+0.19}_{-0.26}$ & $-0.54^{+0.11}_{-0.12}$	\vspace{0.5mm} \\
    $D_{Y_X}$	& $0.147\pm 0.070$	& $0.168^{+0.093}_{-0.064}$	& $0.152^{+0.093}_{-0.078}$	& $0.171^{+0.099}_{-0.058}$	& $0.151^{+0.084}_{-0.073}$ & $0.165^{+0.10}_{-0.066}$	\vspace{0.5mm} \\
    \hline \vspace{-3mm}\\
    \end{tabular}
\end{table*}


\begin{figure*}
\begin{center}
    \includegraphics[width= \textwidth]{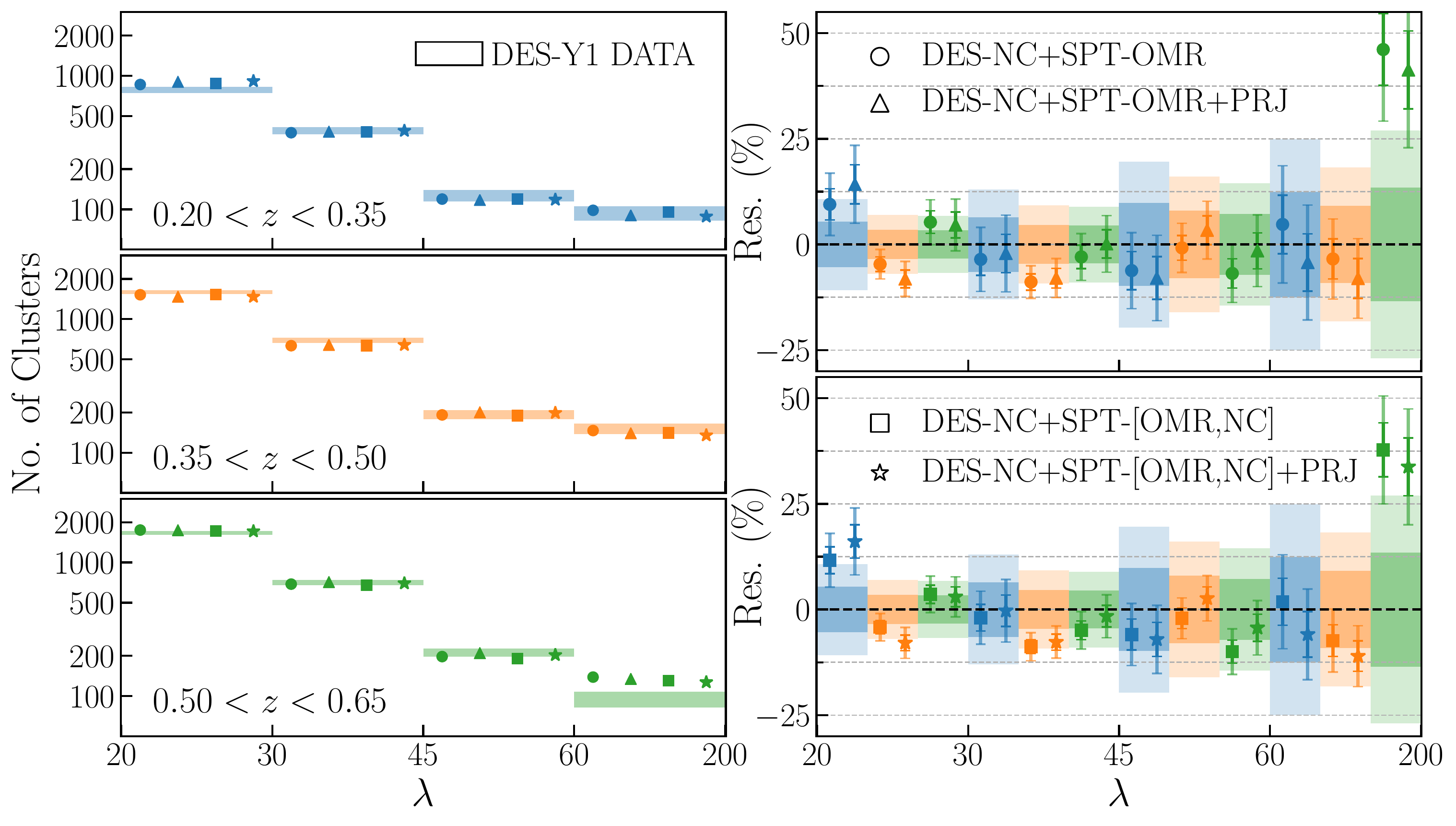}
\end{center}
\caption{Observed (\textit{shaded areas}) and mean model predictions (\textit{markers}) for the DES Y1 RM cluster number counts as a function of richness for each of our three redshift bins. The \textit{y} extent of the data boxes is given by the square root of the diagonal terms of the covariance matrix. The right panels show the residual between the data and the mean model predictions. The error bars on the predicted number counts represent one and two standard deviations of the distribution derived sampling the corresponding chain. All points have been slightly displaced along the richness axis to avoid overcrowding.}
\label{fig:datavec}
\end{figure*}


\section{Results}
\label{sec:res}
Table \ref{tab:results} summarizes the results obtained for the different models and data combinations considered in this work. Along with the varied ones we also report posteriors for two derived parameters: the amplitude of the matter power spectrum on a $8 \hMpc$ scale, $\sigma_8$, and the cluster normalization condition, $S_8=\sigma_8 (\Omega_m/0.3)^{0.5}$.

\subsection{$\Lambda$CDM+$\sum m_\nu$ cosmology}
\label{sec:resLCDM}
Figure~\ref{fig:LCDM_constraints} shows the parameter posteriors obtained from the four analyses carried out for the $\Lambda$CDM+$\sum m_\nu$ model. We do not report posteriors for those parameters not constrained by our data or dominated by their priors. Also, to avoid overcrowding we omit from this figure the $Y_X$ scaling relation parameters which can be found in appendix \ref{app:yx} along with the correlation matrix for a sub-set of parameters. 
The only two cosmological parameters constrained by our data are $\Omegam$ and $\sigma_8$. For all the other cosmological parameters --- $\Omegab h^2$, $\Omega_\nu h^2$ and $n_s$ --- we obtain almost flat posteriors, but for the Hubble parameter which is loosely constrained by the abundance data thanks to the mild sensitivity of the slope of the halo mass function and comoving volume element to variation of $h$.

\subsubsection{Models and data combinations comparison}
\label{sec:res_models}
The left panels of Figure~\ref{fig:datavec} compare the abundances of the DES Y1 RM clusters (\textit{boxes}) with the corresponding mean model predictions (\textit{markers}). The right panels show the residuals between the data and the model expectations for the two scatter models and data combinations considered. 

Starting with our baseline data set DES-NC+SPT-OMR, the SPT multi-wavelength data carry the information to constrain the observable--mass relation parameters, while the DES Y1 RM abundance data, thanks to the SPT-OMR calibrated richness--mass relation, constrain the cosmological parameters. Specifically, the richness--mass relation parameters are constrained through the calibration of the $\xi$--mass scaling relation, which in turn is primarily informed by the weak lensing data. The X-ray data mainly affect the constraints on the intrinsic scatter parameters \citep[see also][]{Bocquet2018}. We explicitly verified that when dropping the X-ray data, we obtain perfectly consistent results for all parameters but for the scatters $D_{SZ}$ and $D_{\lambda}$ whose mean values increase and decrease by $\sim 0.1$ ($\sim 1\sigma$), respectively. 

The further inclusion of the SPT-NC data bring additional cosmological information which slightly improves the $\sigma_8$ and $\Omegam$ constraints --- by $30\%$ and $20\%$, respectively --- while shifting their confidence contours along the $S_8$ degeneracy direction (\textit{black dashed} and \textit{green} contours in figure \ref{fig:LCDM_constraints}). The shift of the $\sigma_8$ posterior can be understood by looking at figure~\ref{fig:sptnc} which compares the SPT-SZ number count data with predictions from the DES-NC+SPT-OMR and DES-NC+SPT-[OMR,NC] analyses. The larger $\sigma_8$ value preferred by the DES-NC+SPT-OMR data tend to over-predict the number of SPT-SZ clusters above $z>0.65$. Consequently, when included, the SPT number count data shift $\sigma_8$ towards lower values to recover the correct number of SPT-SZ clusters (see also \textit{orange} contours in figure \ref{fig:s8_om_comp}). 
Concurrently, to counterbalance the lower $\sigma_8$ mean value and thus keep roughly unvaried the predictions for the DES Y1 RM cluster counts, $\Omega_m$, $B_\lambda$ and $C_\lambda$ move toward larger values following the corresponding degeneracy directions with $\sigma_8$. 
We will further comment on the origin of this shift in section \ref{sec:scal_rel}.
Finally, the SPT abundance data improve the constraints on $B_\sz$ and $C_\sz$ thanks to the sensitivity of the SPT-NC likelihood to the SZ--mass scaling relation.

Moving to the modeling of $P(\lob|\lambda)$, we find consistent results between the two models adopted for the observational noise on $\lambda$ (BKG with \textit{orange} and \textit{black} contours and PRJ with \textit{blue} and \textit{green} contours; see section \ref{sec:mor}), albeit the PRJ model prefers a slightly lower $\Omegam$ value, driven by a shallower ($B_\lambda = 0.86 \pm 0.04$) and redshift independent ($C_\lambda = -0.02 \pm 0.34 $) richness--mass scaling relation compared to the BKG results. This result can be understood as follows: the PRJ model, which accounts also for projection and masking effects, tends to bias high the richness estimates and introduces a larger scatter between $\lob$ and $\lambda$ compared to the BKG model.
As a consequence, for a given set of cosmological and scaling relation parameters, the slope of the $\lob$--mass relation increases, as well as the predicted cluster counts for DES Y1 RM. Given the strong degeneracy between $A_\lambda-A_\sz$ and $D_\lambda-D_\sz$, and the tight constraints on SZ parameters provided by the SPT-OMR data, 
$B_\lambda$ is the only parameter which can compensate for such effects by moving its posterior to lower values.
Similarly, the preference for a non-evolving $\lambda$--mass scaling relation is explained by the redshift dependent bias and scatter intrinsic to the PRJ model, which is a consequence of the worsening of the photo-z accuracy with increasing redshift.
These findings are consistent with those obtained in \citetalias{desy1cl}, where it is shown the robustness of the cosmological posteriors to different model assumptions for $P(\lob|\lambda)$.

As for the correlation coefficients between scatters in all the four cases analyzed the posteriors are prior dominated. We note, however, that while the posteriors of the correlation coefficients between SZ and WL and WL and $\lambda$ peak around zero, the $\rho(\sz,\lambda)$ posterior always has its maximum at $\sim -0.2$, suggesting an anti-correlation between the two observables (see figure \ref{fig:LCDM_constraints_all} in appendix \ref{app:yx}).

\subsubsection{Goodness of fit}
\label{sec:goodness}

The four analyses perform similarly well in fitting the DES Y1 abundance data.
The model predictions are all consistent within $2\sigma$ with the data but for the highest richness/redshift bin, where all the models over-predict the number counts by $\sim 35 \%$ (see \textit{right} panels of figure~\ref{fig:datavec}).
Notably, while the SPT-OMR data is only available for clusters above $\lob \gtrsim 40$, the scaling relation extrapolated at low richness provides a good fit to the DES Y1 abundance data.
Our composite likelihood model and parameter degeneracies do not allow us to apply a $\chi^2$ statistic to assess the goodness of the fit. 
The same tensions between predictions and DES Y1 RM abundance data was observed in \citetalias{desy1cl}, where the authors verified that dropping the highest-$\lambda/z$ bin from the data does not affect their results, but improve the goodness of the fit.
Here we use the posterior predictive distribution to asses the likelihood of observing the highest-$\lambda/z$ data point given our models \citep[see e.g.][section 6.3]{gelmanbda04}. The method consists of drawing simulated values
from the posterior predictive distribution of replicated data and comparing these mock samples to the observed data. The posterior predictive distribution is defined as:
\begin{equation}
    \label{eqn:ppd}
    P(y^{\rm rep}|y)=\int \de \theta P(y^{\rm rep}|\theta) P(\theta|y)
\end{equation}
where $y$ is the observed data vector, $y^{\rm rep}$ the replicated one, and $\theta$ the model parameters. In practice, we generate our replicated data for the highest-$\lambda/z$ by sampling the posterior distribution, $P(\theta|y)$, and drawing for each sampled $\theta$ a value from the multivariate normal distribution defined by equation \ref{eqn:nc} and covariance matrix ${\bm C}$.
We draw $500$ samples for each of the four analyses, and fit the distributions with a Gaussian to easily quantify the likelihood of the observed data point. As can been seen in figure \ref{fig:ppd} for the two models and data combinations considered here the observed data lies within the $3\sigma$ region (\textit{dashed} and \textit{dotted} vertical lines), thus we conclude that the highest-$\lambda/z$ data point is not a strong outlier of the predicted distribution and our model suffices to describe it.

\begin{figure}
\begin{center}
    \includegraphics[width=0.45 \textwidth]{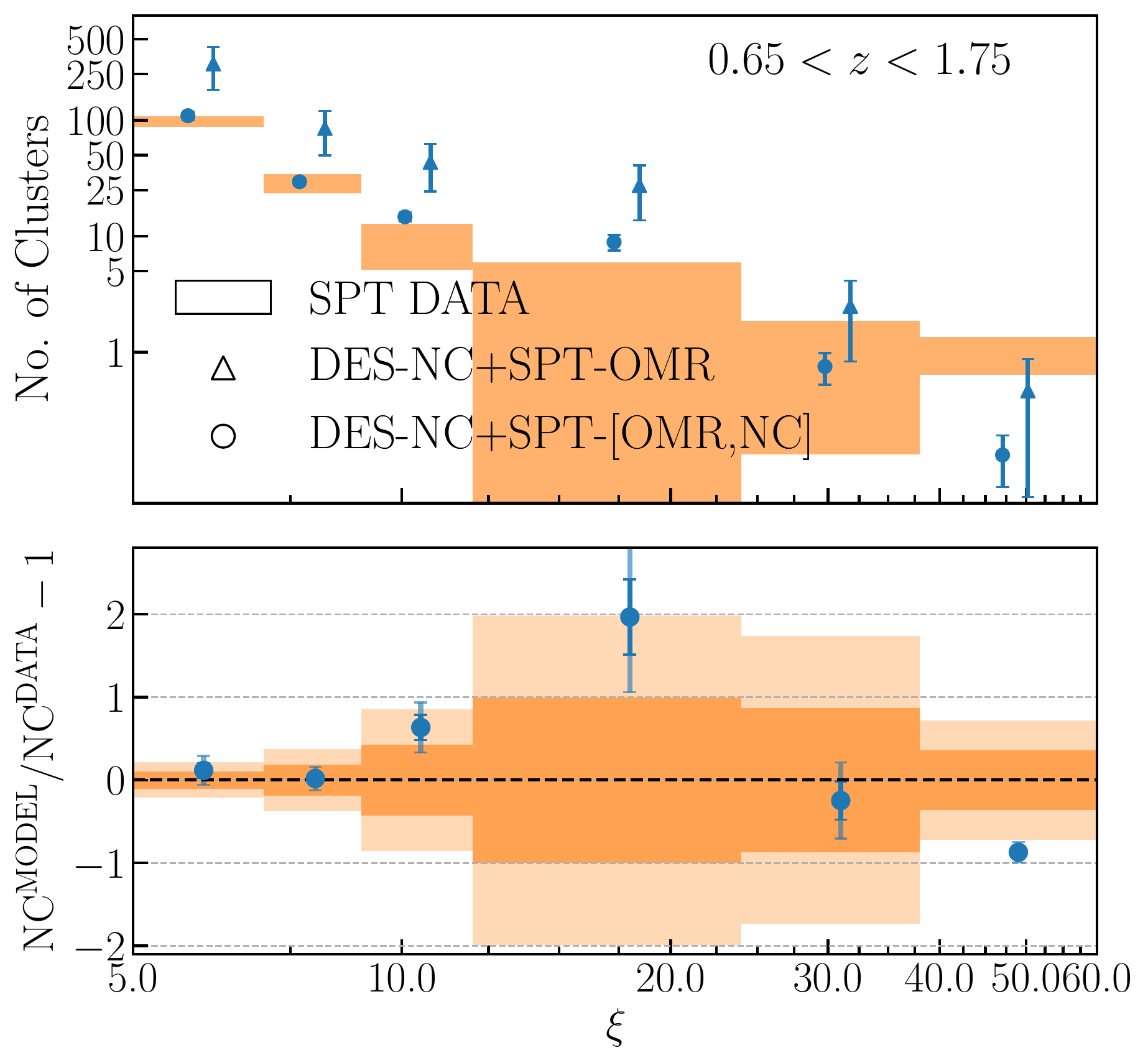}
\end{center}
\caption{Observed SPT-SZ cluster number counts (\textit{shaded areas}) and mean model predictions from the DES-NC+SPT-OMR (\textit{triangles}) and DES-NC+SPT-[OMR,NC] (\textit{circles}) analyses, as a function of $\xi$. The points have been slightly displaced along the $\xi$ axis to avoid overcrowding. The \textit{y} extent of the data boxes corresponds to the Poisson noise. The bottom panel shows the residual between the data and the mean model predictions derived from DES-NC+SPT-[OMR,NC]. The error bars on the predicted number counts represent one and two standard deviations of the distribution derived sampling the relevant parameters of the corresponding chain. The \textit{y} extent of the data boxes corresponds to one and two standard deviations of the associated Poisson distribution.
The SPT-NC model predictions for the two analyses including the PRJ model are fully consistent with those obtained from the baseline model, and thus not included in the plot to avoid overcrowding.}
\label{fig:sptnc}
\end{figure}

\begin{figure}
\begin{center}
    \includegraphics[width=0.47 \textwidth]{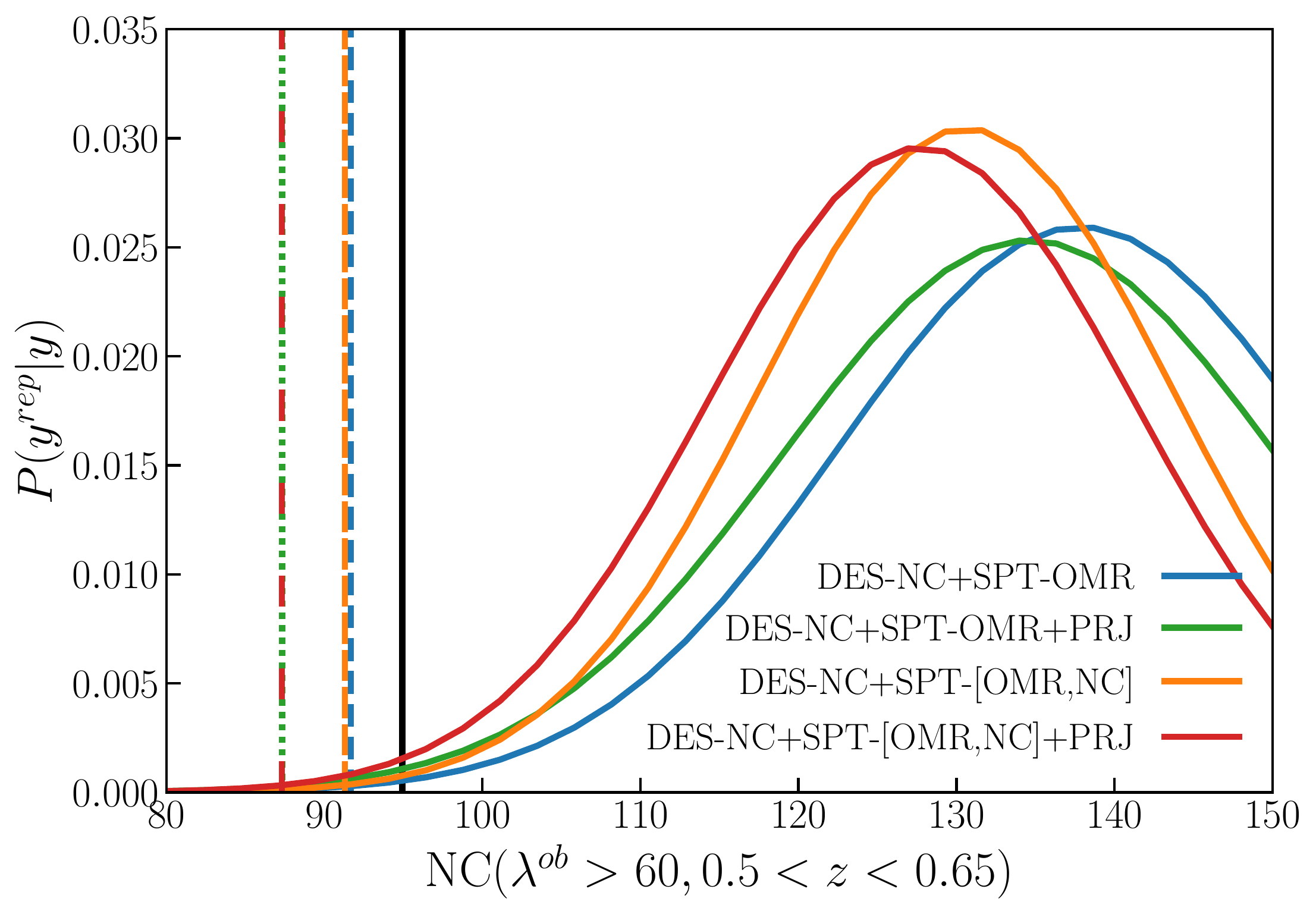}
\end{center}
\caption{Posterior predictive distributions for the highest-$\lambda/z$ data point derived from the four analyses considered in section \ref{sec:resLCDM}. The \textit{solid black} line correspond to the observed cluster abundance in that bin, while the four \textit{dashed} and \textit{dot-dashed} lines mark the 3$\sigma$ limit of the corresponding posterior predictive distribution. Although residing in the tail of the distributions, in none of the four analyses the observed data point lies outside the 3$\sigma$ region. }
\label{fig:ppd}
\end{figure}

\begin{figure*}
\begin{center}
    \includegraphics[width=0.45 \textwidth]{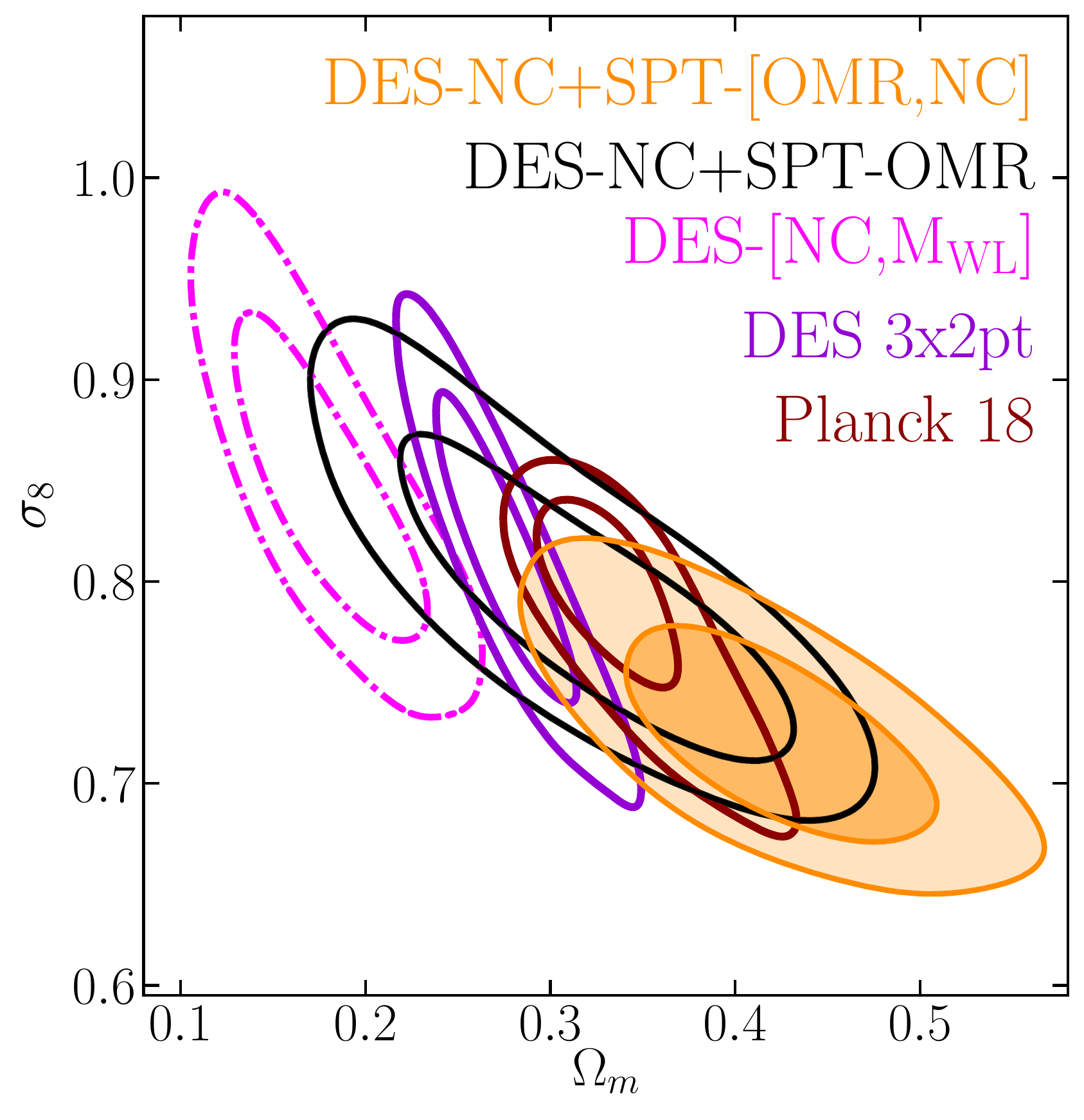}
    \includegraphics[width=0.455 \textwidth]{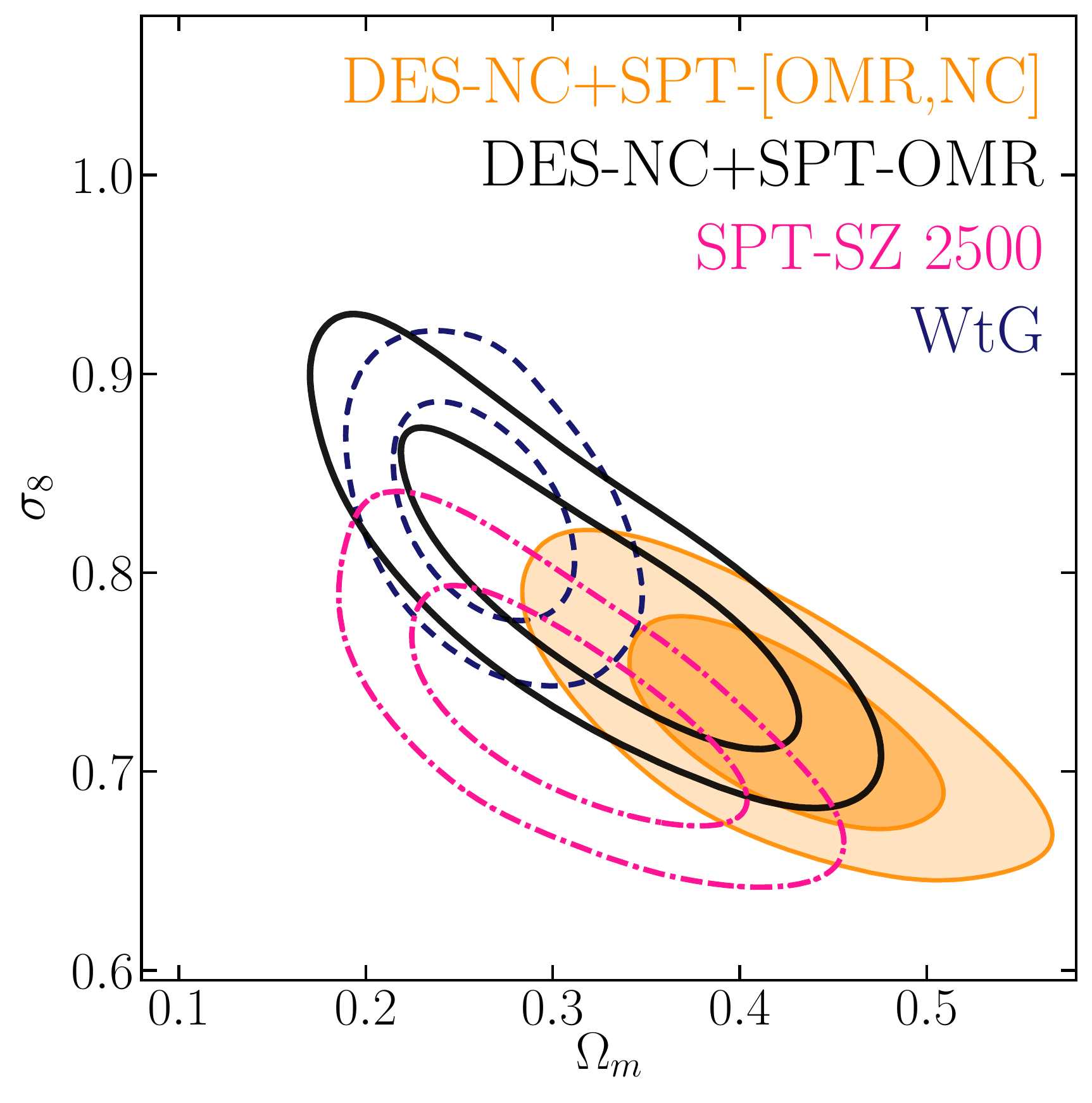}
    \includegraphics[width=0.45 \textwidth]{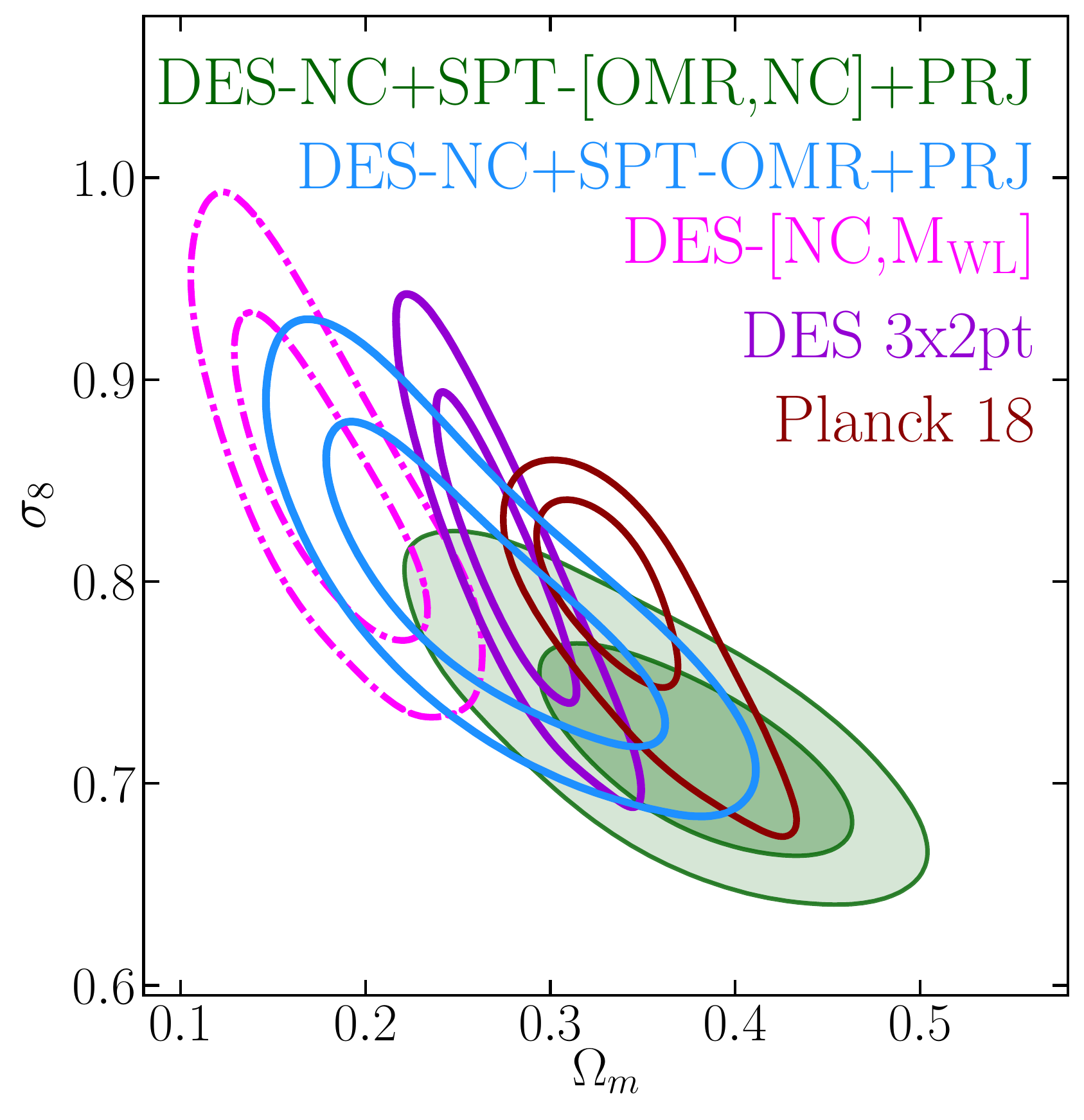}
    \includegraphics[width=0.45 \textwidth]{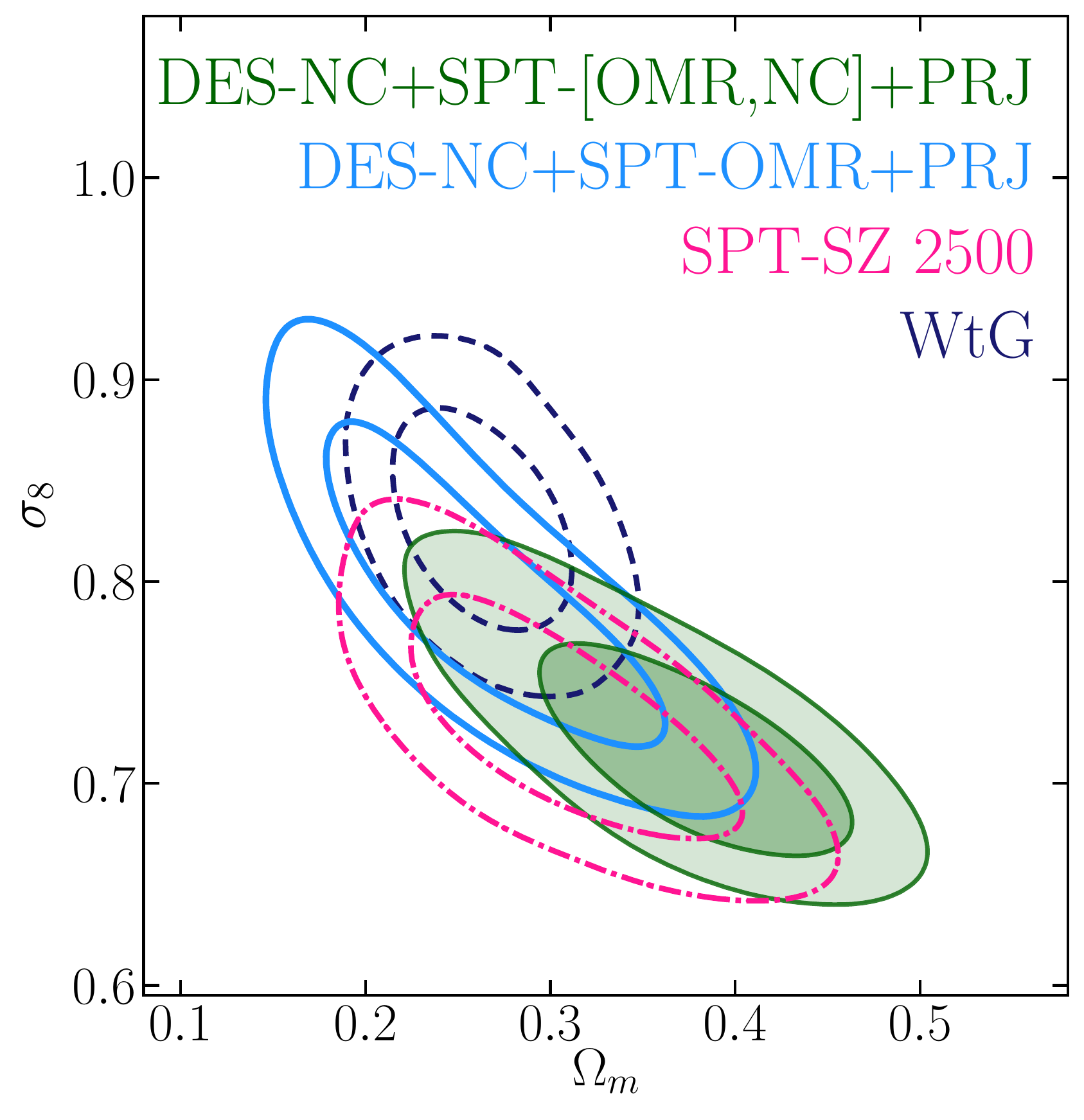}
\end{center}
\caption{\textit{Upper} panels: Comparison of the $68\%$ and $95\%$ confidence contours in the $\sigma_8$-$\Omega_m$ plane derived in this work adopting the BKG scatter model (\textit{black} and \textit{orange} contours) with other constraints from the literature: DES Y1 cluster counts and weak lensing mass calibration \citepalias[][\textit{dot--dashed magenta} contours]{desy1cl}; DES-Y1 3x2 from \cite[][\textit{dark violet} contours]{des17}; \Planck\ CMB from \cite[][\textit{brown} contours]{Planck2018}; cluster number counts and follow-up data from the SPT-SZ 2500 survey \citepalias[][\textit{dot-dashed pink} contours]{Bocquet2018}; cluster abundance analysis of Weighing the Giants \citep[][WtG, \textit{dashed dark blue} contours]{Mantz2015}. \textit{Lower} panels: Same as \textit{left} panel but considering the projection effect model (PRJ) for the scatter between true and observed richness (see section \ref{sec:mor}).
}
\label{fig:s8_om_comp}
\end{figure*}

Similarly for the SPT-SZ abundance data, the models retrieved from the posteriors of the DES-NC+SPT-[OMR,NC] and DES-NC+SPT-[OMR,NC]+PRJ analyses provide a good fit to the SPT number counts but for the highest $\xi$ bin, where the model predictions lie at the edge of the $\sim 2\sigma$ region (see lower panel of figure \ref{fig:sptnc}). 

As for the SPT-OMR data we inspect the goodness of the fit of the derived $P(\lob|\xi)$ distributions against the cross-matched sample. Specifically, we verified that all the data points lie within the $3\sigma$ region of the posterior predictive distributions independently from the data combination and model assumed for the observational scatter on $\lob$ (see figure \ref{fig:lob_xi_rel}).

To determine whether our data sets prefer one of the two models adopted for $P(\lob|\lambda)$ --- BKG and PRJ --- we rely on the Deviance Information Criterion \citep[hereafter DIC;][]{DIC2002}. Specifically, for a given model $M$ the DIC is computed from the mean $\chi^2$ over the posterior volume and the maximum posterior $\chi^2$ as:
\begin{equation}
\label{eqn:dic}
    {\rm DIC}(M)=2 \langle \chi^2 \rangle_M -\chi^2_{\rm MaxP}(M) \, .
\end{equation}
The model with the lower {\rm DIC} value either fits better the data --- lower $\langle \chi^2 \rangle$ --- or has a lower level of complexity --- lower $\left( \langle \chi^2 \rangle -\chi^2_{\rm MaxP}\right)$. For the data combination DES-NC+SPT-OMR we obtain $\Delta {\rm DIC} = {\rm DIC}(\rm PRJ) - {\rm DIC}(\rm BKG)=3.5$, while for the full data set $\Delta {\rm DIC} =-3.8$. Adopting the Jeffreys' scale to interpret the $\Delta {\rm DIC}$ values, the DES-NC+SPT-OMR data combination has a "positive" ($|\Delta {\rm DIC}| \in [2, 5]$) -- even though  not "strong" ($|\Delta {\rm DIC}| \in [-5, -10]$) or "definitive" ($|\Delta {\rm DIC}| > 10$) -- preference for the BKG model, while the full data combination has a "positive" preference for the PRJ model. 
Additional follow-up data extending to lower richness --- as the one soon available from the combination of DES Y3 and Y6 data with the full SPT surveys or eROSITA --- will help to identify the model which better describes the data.
 
%
%
%
%

%
%

\subsubsection{Comparison with other cosmological probes}
\label{sec:comparison}
Figure \ref{fig:s8_om_comp} compares the $\sigma_8$-$\Omega_m$ posteriors derived in this work for a \LCDM+$\sum m_\nu$ cosmology including (\textit{lower} panels) or excluding (\textit{upper} panels) the PRJ calibration, to other results from the literature. To assess the consistency of two data sets $A$ and $B$ in the $\sigma_8$-$\Omega_m$ plane we test the hypothesis $\bm{p}_A - \bm{p}_B = 0$  \citep[see method `3' in][]{charnocketal17},  where $\bm{p}_A$ and $\bm{p}_B$ are the $\sigma_8$-$\Omega_m$ posterior distributions as constrained by data sets $A$ and $B$, respectively.

Starting with the simpler scatter model (BKG, \textit{upper} panels), our baseline data combination (DES-NC+SPT-OMR) is consistent within $2\sigma$ with all the probes considered here.
The largest tension ($1.7\sigma$) is found with the results from \citetalias{desy1cl} (DES-[NC,$\mwl$] in figure \ref{fig:s8_om_comp}) which combine DES Y1 RM abundances and mass estimates from the stacked weak lensing signal around DES Y1 RM clusters \citep{desy1wl}. The tension with DES-[NC,$\mwl$] results is not surprising and reflects the different richness--mass scaling relation preferred by the DES Y1 weak lensing calibration (see also section \ref{sec:scal_rel}). The consistency of our posteriors with the DES Y1 combined analysis of galaxy clustering and weak lensing \citep[DES 3x2pt][]{des17}, \planck\ CMB data \citep{Planck2018}, and other cluster abundance studies, seems to confirm the conclusions of \citetalias{desy1cl}: the tension between DES-[NC,$\mwl$] and other probes is most likely due to flawed interpretation of the stacked weak lensing signal of \redmapper\ clusters in terms of mean cluster mass.

The similar constraining power provided by our data set and SPT-SZ 2500, which combine SPT-OMR data and SPT-SZ cluster counts above $z>0.25$, indicates that the two analyses are limited by the uncertainty in the mass calibration, i.e. the data set they have in common.
The lower $\sigma_8$ value preferred by the SPT SZ-2500 analysis \citep{Bocquet2018}\footnote{Note that at odds with the \citetalias{Bocquet2018} analysis, here we show results for the SPT-SZ 2500 analysis obtained assuming 3 degenerate massive neutrino species and adopting the massive neutrino prescription for the halo mass function presented in \cite{Costanzi2013}, consistently with our analysis. The different massive neutrino scheme and the inclusion of this prescription lowers the $\sigma_8$ posterior by $0.024$ (corresponding to $\sim 0.5\sigma$) compared to original results of \citetalias{Bocquet2018}.} can be again understood by looking at figure~\ref{fig:sptnc}: the cosmology preferred by the DES-NC+SPT-OMR data combination over-predict the SPT-NC by a factor of $\sim 2$, and the same trend holds for the SPT abundance data below $z=0.65$ (not shown in the figure). As a consequence, when substituting the DES-NC data with the SPT-SZ cluster number counts, the $\sigma_8$ posterior shifts toward lower values to accommodate the model predictions to the new abundance data.

The inclusion of SPT-NC data (DES-NC+SPT-[OMR,NC]) worsens the consistency with the other low-redshift probes considered here by shifting the $\Omegam$/$\sigma_8$ posteriors towards higher/lower values. In particular, the agreement is degraded with the DES 3x2pt and WtG results, with which the tension in the $\sigma_8$-$\Omega_m$ plane raises to $1.8\sigma$ and $1.9\sigma$, respectively.
Notably, the full data combination is at $1.3\sigma$ tension also with results from SPT-SZ 2500 with which it shares part of the abundance data (SPT-SZ counts above $z=0.65$) and the follow-up data.
The fact that the DES-NC+SPT-[OMR,NC] posteriors do not lie in the intersection of the DES-NC+SPT-OMR and SPT-SZ 2500 contours suggests the presence of some --- yet not statistically significant --- tension between the DES-NC, SPT-OMR and SPT-NC data, possibly driven by an imperfect modeling of the scaling relations\footnote{To exclude the possibility that the tension is driven by SPT-SZ abundance data at low redshift we re-analyze the SPT-SZ 2500 catalog excluding the cluster counts data below $z=0.65$ --- i.e. analysing the data combination SPT-[OMR,NC] --- finding posteriors fully consistent with SPT-SZ 2500 results \citep[see also figure 16 in][]{Bocquet2018}.}.

On the other hand, by turning the $\sigma_8$-$\Omega_m$ degeneracy direction, the inclusion of the PRJ model (\textit{lower} panel) improves the agreement of the DES-NC+SPT-OMR posteriors with the SPT-SZ 2500 results (from $1\sigma$ to $0.5\sigma$ tension), at the expense of larger, yet not significant ($1.3\sigma $), tension with CMB data (\textit{red} contours).
Also the tension with the \citetalias{desy1cl} results decreases ($0.7\sigma$) as a consequence of the improved consistency between the richness--mass scaling relations (see section \ref{sec:scal_rel}).
Similarly, when considering the full data combination, the PRJ model shifts the cosmological posteriors in the intersection of the DES-NC+SPT-OMR  and SPT-SZ 2500 contours, solving the above mentioned tension between the three data set. We will go back to this point in the next section.


\subsubsection{The mass--richness relation}
\label{sec:scal_rel}
Being constrained by the SPT multi-wavelength data both the SZ and $Y_X$ scaling relations derived from the DES-NC+SPT-OMR analysis are perfectly consistent with those obtained in \citetalias{Bocquet2018}. The inclusion of SPT-NC data in our analysis shifts the slope of the SZ relation, $B_\sz$, by $1.5\sigma$ towards steeper values to compensate for the larger $\Omegam$ value preferred by the full data combination. As mentioned before, the shift of the cosmological posteriors along the $S_8$ direction suggests the presence of some inconsistencies between the scaling relations preferred by the different data sets: DES-NC, SPT-OMR and SPT-NC. To pinpoint the source of tension we re-analyze the abundance and multi-wavelengths data independently using as cosmological priors the product of the posterior distributions obtained from the DES-NC+SPT-OMR and SPT-[OMR,NC] analyses (roughly the intersection between the \textit{black} and \textit{pink} contours in the upper right panel of figure~\ref{fig:s8_om_comp}). This test will allow us to understand why that region of the $\sigma_8$--$\Omegam$ plane is disfavored by the full data combination.

\begin{figure}
\begin{center}
    \includegraphics[width=0.47\textwidth]{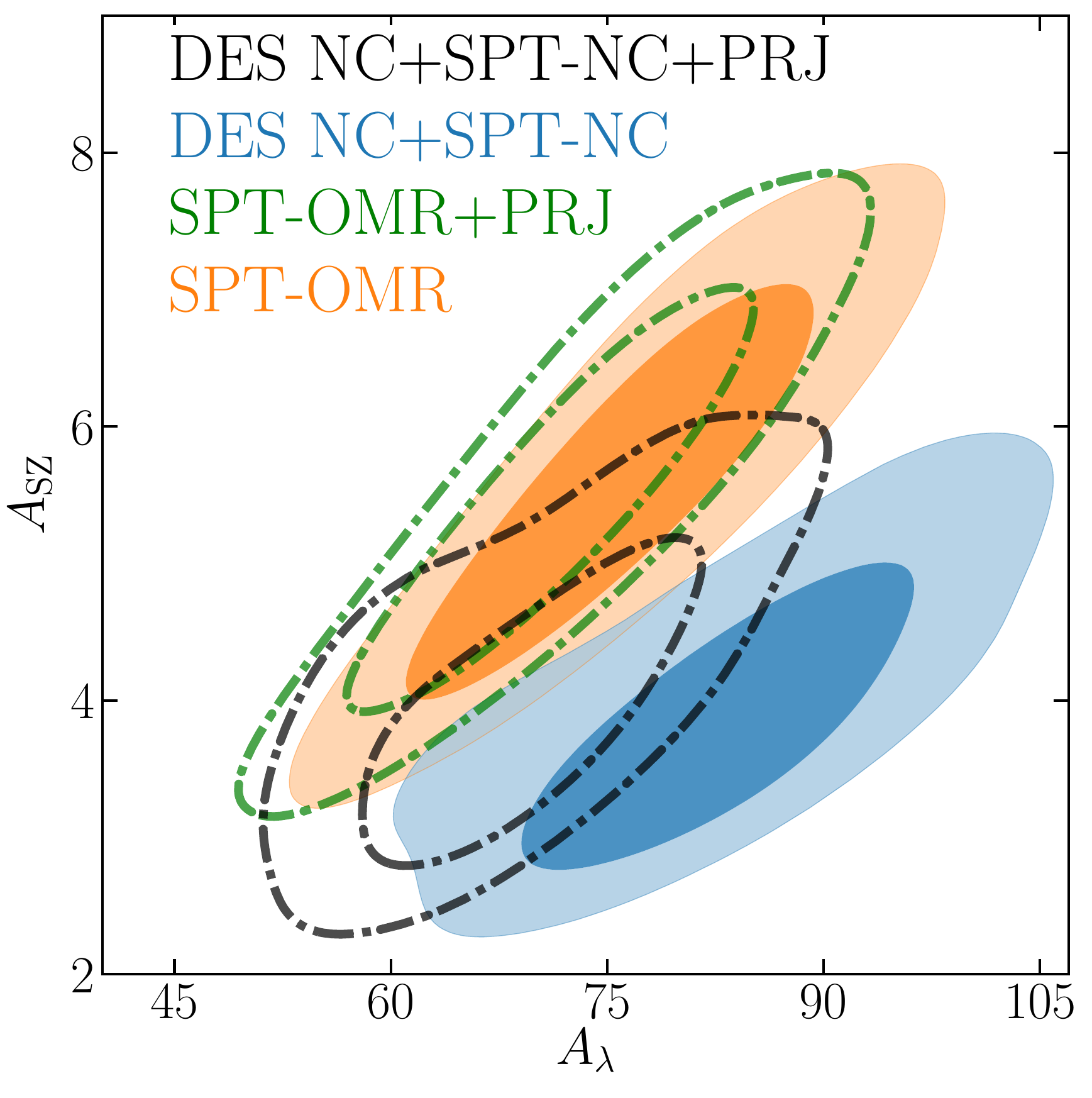}
\end{center}
\caption{$68\%$ and $95\%$ confidence contours for the amplitude parameters $A_\lambda$-$A_\sz$ from the combination of DES Y1 and SPT cluster counts data (\textit{blue} and \textit{black}) or the SPT multi-wavelength data (\textit{orange} and \textit{green}), including (\textit{dot-dashed} contours) or not (\textit{filled} contours) the projection effect model (PRJ). All the contours are derived imposing the cosmological priors resulting from the combination of the posteriors obtained from the DES-NC+SPT-OMR and SPT-[OMR,NC] analyses. By shifting the abundance posteriors towards lower $A_\lambda$ values (\textit{black} versus \textit{blue} contours) the PRJ model relieves the tension between the scaling relation parameters preferred by abundance and multi-wavelength data.}
\label{fig:SZ_vs_lam}
\end{figure}

As can been seen in figure~\ref{fig:SZ_vs_lam} the tension between DES-NC+SPT-OMR, SPT-NC and DES-NC+SPT-[OMR,NC] arises from the different amplitude of the richness and SZ scaling relation preferred by the abundance (\textit{blue} contours) and SPT-OMR data (\textit{orange} contours). The PRJ model, lowering the $A_\lambda$ value preferred by the abundance data (\textit{black dot-dashed} contours), but leaving almost unaffected the SPT-OMR posteriors (\textit{green dot-dashed} contours), largely alleviates the tension between data sets.
Once we let the cosmological parameters free to vary, the tight correlation between the SZ and richness scaling relation parameters introduced by the SPT-OMR data, along with the different posteriors for the amplitudes preferred by the latter, moves the $\Omegam$ posterior of the full data combination towards larger values.
The larger shift with respect to the DES-NC+SPT-OMR data combination observed for the BKG analysis can be understood in terms of the larger tension between multi-wavelength and abundance data displayed in figure ~\ref{fig:SZ_vs_lam}.
Despite the better agreement of the $A_\lambda$-$A_{\rm SZ}$ posteriors derived assuming the PRJ calibration, the DIC suggests a mild preference for this model only for the full data combination (see section \ref{sec:res_models}).

\begin{figure}
\begin{center}
    \includegraphics[width=0.47 \textwidth]{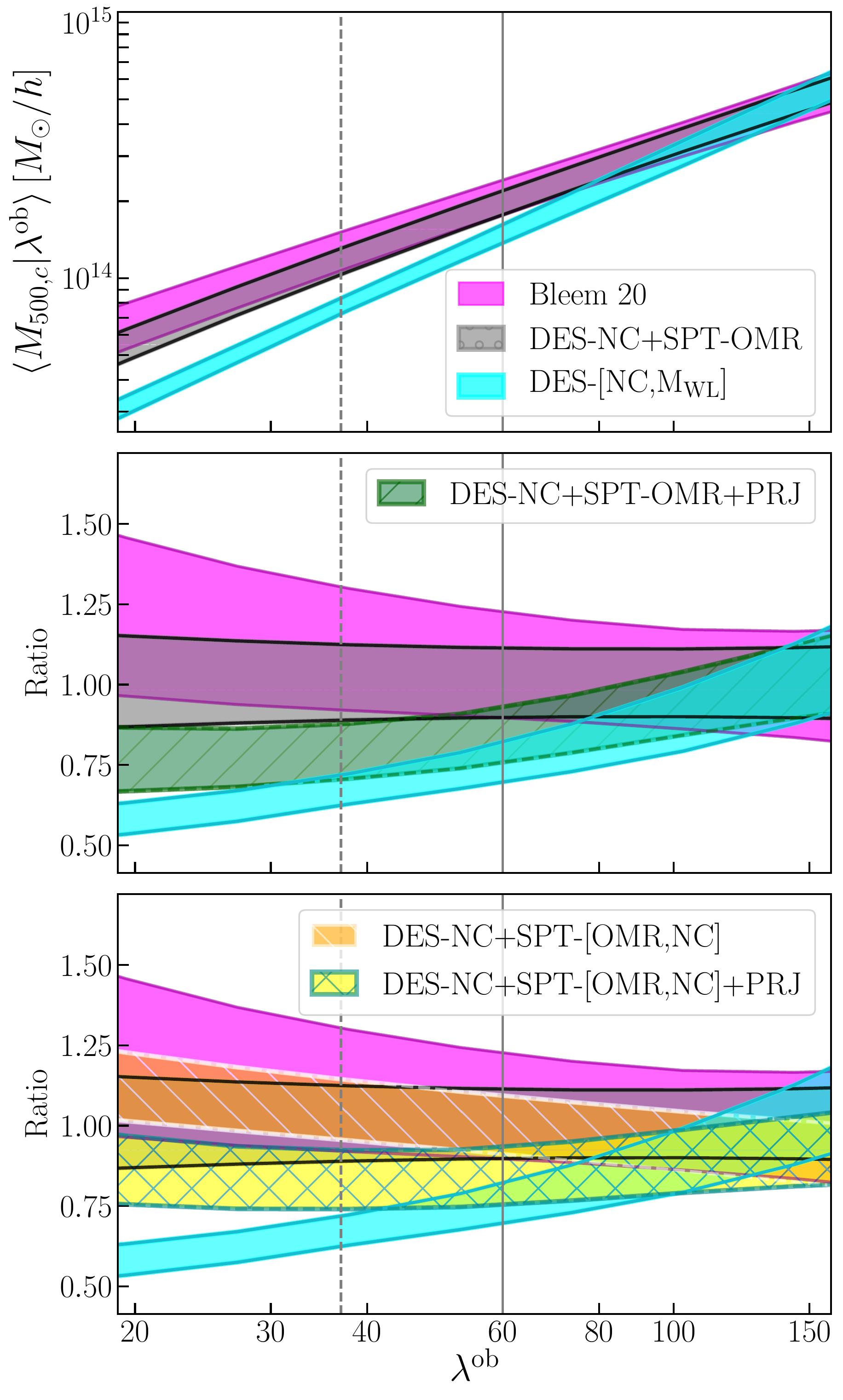}
\end{center}
\caption{Comparison of mass--richness relations at the mean DES Y1 RM redshift $z=0.45$. The \textit{gray}, \textit{green}, \textit{orange} and \textit{yellow} bands show the M-$\lambda\ob$ relations derived in this work for different models and data combinations. Shown in \textit{magenta} is the $\langle M | \lambda\ob \rangle$ relation derived by \cite{Bleem2020} using SPT SZ cluster counts and follow-up data, assuming a \planck\ cosmology. The relation derived in \citetalias{desy1cl} combining DES Y1 number counts and weak lensing mass estimates is shown with the \textit{cyan} band. The $y$ extent of the bands corresponds to $1\sigma$ uncertainty of the mean relation. The \textit{lower} panels show the ratio of the different mass--richness relations to the one derived from the DES-NC+SPT-OMR analysis. The \textit{dashed} ($\lambda\ob=37$) and \textit{solid} ($\lambda\ob=60$) vertical lines correspond to the richnesses above which $95\%$ and $68\%$ of the DES Y1 RM-SPT-SZ matched sample is contained.}
\label{fig:M_lob_rel}
\end{figure}

Moving to the mass--richness relation, figure \ref{fig:M_lob_rel} compares the scaling relations derived in this work (\textit{hatched} bands) with other results from the literature.
The scaling relation from \citetalias{desy1cl} originally derived for $M_{200,m}$ has been converted to $\langle M_{500,c}|\lob,z \rangle$ imposing the condition $n(M_{500,c})\de M_{500,c} = n(M_{200,m}) \de M_{200,m}$ to the Tinker halo mass function.
The mean mass-richness relation and its uncertainty are computed from the $\lambda$--mass parameter posteriors through Bayes' theorem as follows:
\begin{equation}
\label{eqn:M_lam}
\langle M|\lob,z \rangle \propto \int \de M\, \de \lambda\, M \,  n(M,z) \, P(\lob|\lambda,z) P(\lambda|M,z) \, .   
\end{equation}

Fitting the $\langle M|\lob,z \rangle$ relation derived from DES-NC+SPT-OMR to a power law model similar to the one assumed in \cite{desy1wl} we get\footnote{The corresponding mean richness--mass relations, $ \langle \lob| M_{500,c},z \rangle$, for both scatter models are reported for completeness in appendix \ref{app:lm}.}:
\begin{multline}
    \langle M_{500,c}|\lob,z \rangle = 10^{14.29\pm 0.03} \left(\frac{\lob}{60}\right)^{1.11\pm 0.06}  \\  \left(\frac{1+z}{1+0.35}\right)^{-0.55\pm 0.75}. 
\end{multline}

The DES-NC+SPT-OMR and DES-NC+SPT-[OMR,NC] analyses provide mass--richness relations consistent with each other within one standard deviation (\textit{gray} and \textit{hatched orange} bands).
These results are also consistent with a similar analysis performed by \cite{Bleem2020} who calibrate the $\lambda$--mass relation combining cluster counts from both SPT-SZ and SPTpol Extended Cluster Survey, richnesses obtained by matching the SZ sample with the \redmapper\ DES Year 3 catalog, and assuming the fiducial cosmology $\sigma_8=0.8$ and $\Omega_m=0.3$ (\textit{magenta} band). Also here, the slightly steeper M-$\lambda\ob$ relation preferred by our data is due to the different cosmologies preferred by the DES and SPT abundance data. Indeed, when we include the SPT-NC data in our analysis, the $\langle M|\lob,z \rangle$ relation totally overlaps with the results from \cite{Bleem2020} (\textit{hatched orange} and \textit{magenta} bands).
Similarly, \cite{Grandis} derived a richness--mass relation consistent with ours ($A_\lambda=83.3\pm11.2$ and $B_\lambda=1.03\pm0.10$) analysing the same \redmapper-SPT matched sample and adopting as priors the results of \citetalias{Bocquet2018}.
A consistent slope of the mass--richness relation is also found in the work of \cite[][$B_\lambda= 0.99^{+0.06}_{-0.07} \pm 0.04$]{capasso2018}, who calibrate the richness-mass relation of a X-ray selected, optically confirmed cluster sample through galaxy dynamics. However, a direct interpretation of their results in the context of this analysis is not possible due to the different  assumptions on the X-ray scaling relation and the scatter of the richness--mass relation, made in that work.

A larger than $1\sigma$ tension below $\lambda\ob \simeq 60$ is found with the \citetalias{desy1cl} results which base their mass calibration on the stacked weak lensing analysis of \cite{desy1wl} (\textit{cyan} band in figure \ref{fig:M_lob_rel}). As noted in \citetalias{desy1cl}, the weak lensing mass estimates for $\lambda<30$ are responsible for the low values derived for the slope and amplitude of the richness--mass relation compared to the ones preferred by the SPT multi-wavelength data. We stress again here that the SPT-OMR data can actually constrain the richness--mass relation only at $\lob \gtrsim 50$ and the constraints at low richness follow from the power law model assumed for the $\langle \lambda|M \rangle$ relation.

The inclusion of the PRJ calibration, increasing the fraction of low mass clusters boosted to large richnesses, lowers the mean cluster masses compared to the BKG model up to $\sim 25\%$ at $\lambda\ob \lesssim 60$ (compare \textit{green} and \textit{yellow} with \textit{gray} and \textit{orange} bands in figure \ref{fig:M_lob_rel}, respectively). Specifically, from the DES-NC+SPT-OMR+PRJ analysis we obtain:
\begin{multline}
    \langle M_{500,c}|\lob,z \rangle = 10^{14.22\pm 0.03} \left(\frac{\lob}{60}\right)^{1.21\pm0.05} \\ \left(\frac{1+z}{1+0.35}\right)^{-0.50\pm0.65}.
\end{multline}

The improved consistency between the scaling relations derived from the analyses adopting the PRJ calibration and \citetalias{desy1cl} reflects the improved agreement between the corresponding cosmological posteriors due to the lower $\Omegam$ value preferred by the former (see figure \ref{fig:s8_om_comp}).
The fact that the mass-richness relations derived from the two $P(\lob|\ltrue)$ models display a larger than $1\sigma$ tension below $\lob \lesssim 50$, but perform equally well in fitting the data (see section \ref{sec:res_models}), is due to the lack of multi-wavelength data at low richness. Additional follow-up data at $\lob \lesssim 40$ will be fundamental to clearly reject one of the models and thus enable the full exploitation of the cosmological information carried by photometrically selected cluster catalogs.

It is worth noting that at odds with other studies which rely on stacked weak lensing measurements to calibrate the mean scaling relation \citep[e.g.][]{costanzietal18b,desy1cl}, the SPT-OMR data, allowing a cluster by cluster analysis (see equation \ref{eqn:likeMOR}), enable to constrain also the scatter of the richness--mass relation.
This is particularly relevant for the analysis of optically--selected cluster samples for which reliable simulation-based priors on the scatter are not available: if constrained only by the abundance data, the scatter parameter becomes degenerate with $\Omega_m$ and $\sigma_8$, degrading the constraining power of the sample \citep[e.g. see discussion in][]{costanzietal18b}.

\begin{figure}
\begin{center}
    \includegraphics[width=0.45 \textwidth]{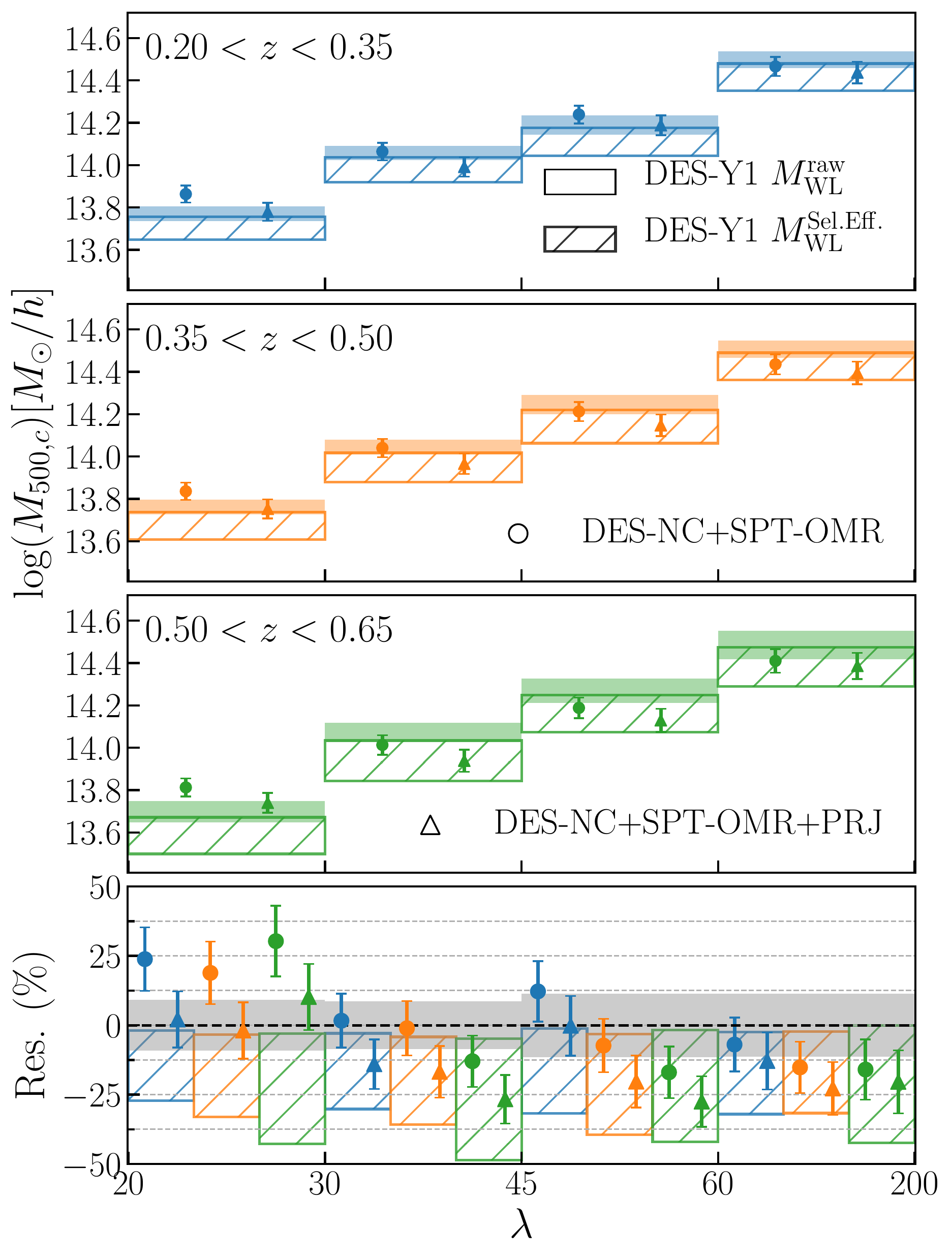}
\end{center}
\caption{Mean mass estimates from the stacked weak lensing analysis of \cite{desy1wl}, including (\textit{hatched boxes}) or not (\textit{filled boxes}) the selection effect bias correction as derived in \cite{Wu}. Over plotted are the mean cluster masses predicted by the scaling relations derived in this work (\textit{circles} and \textit{triangles}). The $y$ extent of the boxes corresponds to uncertainties associated with the mass estimates. The error bars correspond to the $1\sigma$ uncertainty of the models as derived from the corresponding posterior distributions. The model predictions for the analyses including the SPT-NC data are fully consistent with those obtained from the analyses combining DES-NC and SPT-OMR data, and thus not included in the plot to improve the readability. }
\label{fig:Mwl_comp}
\end{figure}

To better investigate the implications of the derived scaling relation for low richness objects we compare in figure \ref{fig:Mwl_comp} our predictions for the mean cluster masses in different richness/redshift bins to the mean weak lensing mass estimates from \cite{desy1wl} (\textit{filled} boxes). We also include the weak lensing mass estimates employed in the DES Y1 cluster analysis (\textit{hatched} boxes) which adopt an updated calibration of the selection bias based on the simulation analysis of \cite{Wu} \citep[see also appendix D of][]{desy1cl}. Both weak lensing mass estimates and mean mass predictions have been derived assuming $\Omegam=0.3$, $h=0.7$ and $\sigma_8=0.8$\footnote{The larger tension seen in figure \ref{fig:M_lob_rel} between the scaling relations derived in this work and \cite{desy1wl} is due to the different cosmology preferred by the two analyses.}.
The mean mass predictions for the DES-NC+SPT-OMR analysis are in tension with both weak lensing mass estimates.
In particular, in the lowest richness bins, $\lob \in [20,30]$, the mean mass predictions are $25\%$ to $40\%$ higher than the weak lensing mass estimates, while they are consistent within 1 $\sigma$ with the lensing masses at $\lob>30$. 
The inclusion of the PRJ model, lowering the mean mass predictions, largely reduce the tension at low richness with both weak lensing mass estimates, while at $\lob>30$ the model predictions are consistent within 1 $\sigma$ with the weak lensing masses derived adopting the selection effect bias calibration of \cite{Wu}.
These results are consistent with those of \citetalias{desy1cl}: for the DES Y1 cluster cosmology analysis to be consistent with other probes the weak lensing mass estimates of $\lob<30$ systems need to be boosted.  Or conversely, the weak lensing mass estimates of $\lob<30$ systems are biased low compared to the mean masses predicted by DES Y1 abundance data alone assuming a cosmology consistent with other probes.
As discussed in \citetalias{desy1cl} this tension might be due to an overestimate of the selection effect correction at low richness, or to another systematic not captured by the current synthetic cluster catalogs. 
The good agreement of the PRJ mass predictions with the weak lensing masses adopted in \citetalias{desy1cl} reflects the consistency of our cosmological posteriors with those derived in DES Y1 cluster analysis (see the lower left panel of figure \ref{fig:s8_om_comp}).
The same conclusions last also for the full data combination analyses (not shown in figure \ref{fig:Mwl_comp}), which provide model predictions fully consistent with those obtained from the combination of DES abundances and SPT multi-wavelength data.


\begin{figure}
\begin{center}
    \includegraphics[width=0.47\textwidth]{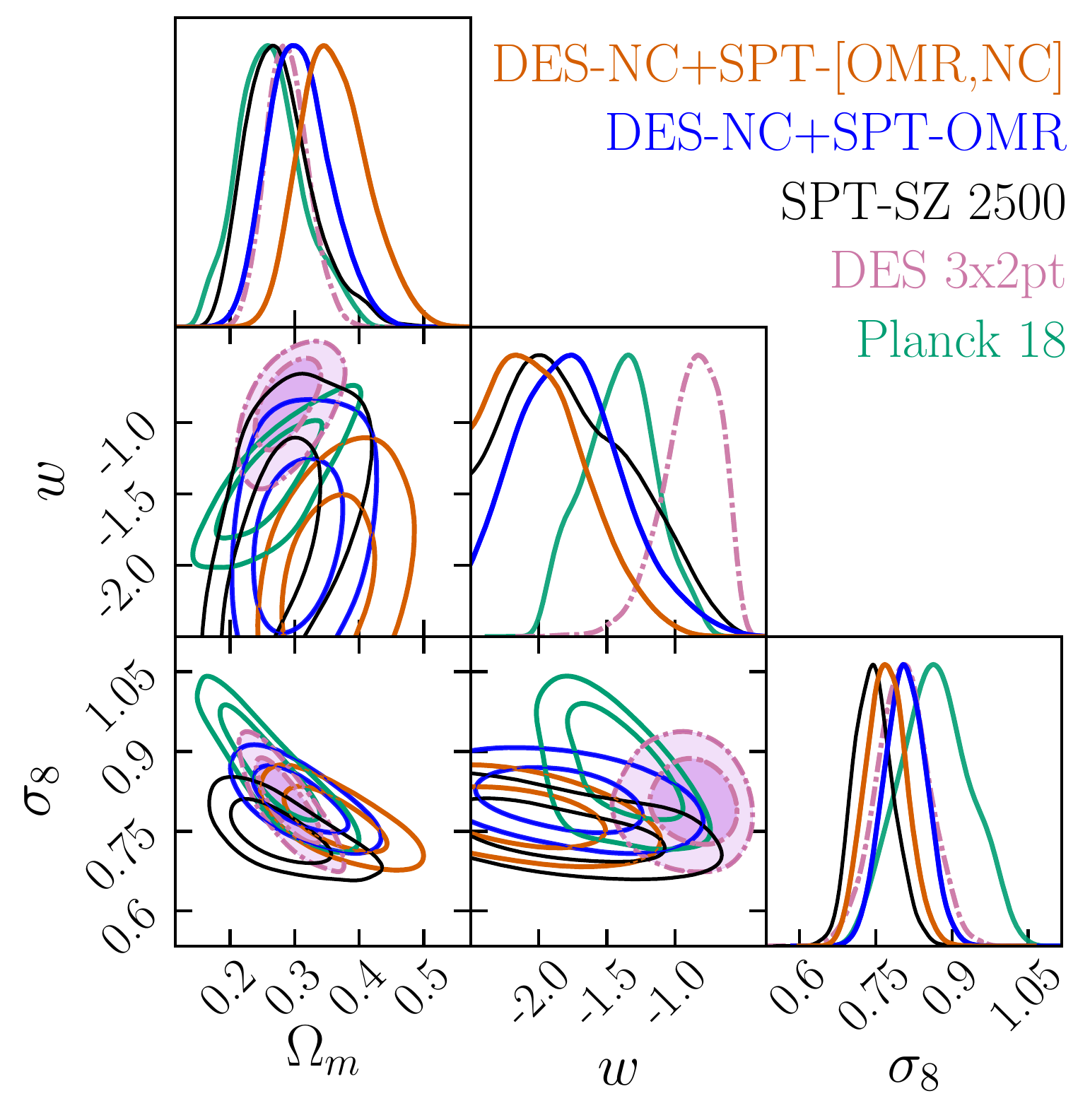}
\end{center}
\caption{Cosmological posteriors ($68\%$ and $95\%$ C.L.) for the $w$CDM+$\sum m_\nu$ model from the combination of DES-NC and SPT-OMR data (\textit{blue}) and the full data combination (\textit{orange}). For comparison we include in the figure the posteriors obtained from \planck\ CMB (\textit{green}), DES 3x2pt (\textit{pink}) and SPT-SZ 2500 (\textit{black}) analyses.}
\label{fig:wCDM}
\end{figure}


\subsection{$w{\rm CDM}+\sum m_\nu$ cosmology}
\label{sec:res_wcdm}
We consider an extension to the vanilla \LCDM model by allowing the dark energy equation of state parameter $w$ to vary in the range $[-2.5,-0.33]$.
Here we are interested in the capability of the DES-NC+SPT-OMR data to constrain the equation of state parameter $w$, and the possible cosmological gain given by the inclusion of the high redshift SPT abundance data. For this reason we report here only results for the BKG scatter model. Nevertheless, we explicitly verified that the PRJ model provides for both data combinations posteriors on $w$ fully consistent with those obtained assuming the BKG model.
In figure \ref{fig:wCDM} and table \ref{tab:results} we show constraints for the DES-NC+SPT-OMR and DES-NC+SPT-[OMR,NC] data sets. Both data sets prefer a $w$ value smaller than -1 at more than one $\sigma$ ($w=-1.76^{+0.46}_{-0.33}$ and $w=-1.95^{+0.48}_{-0.19}$ ), even though consistent within $2\sigma$ with a cosmological constant.
Despite that the inclusion of the SPT-NC data increases the redshift range probed by the abundance data up to $z\simeq 1.75$, the constraints on $w$ improve only by $15\%$.
This again is due to the fact that the analysis is limited by the uncertainty in the calibration of the scaling relations with which the $w$ parameter is degenerate.
For the DES-NC+SPT-OMR analysis the model extension minimally affects the cosmological posteriors on $\sigma_8$ and $\Omega_m$ compared to the \LCDM model despite the mild anti-/correlation of the two parameters with $w$ ($\rho \sim \pm 0.25 $) and the preference for $w<-1$.
Interestingly in this case, the inclusion of the SPT-NC data does not cause the large $\sigma_8$-$\Omega_m$ shift observed in the \LCDM scenario, and the DES-NC+SPT-[OMR,NC] posteriors almost completely overlap with those derived in the DES-NC+SPT-OMR analysis.
This difference with the \LCDM results is explained by the degeneracy of the equation of state parameter $w$ with the SZ and $\lambda$-M scaling relation parameters.
In particular for the DES-NC+SPT-OMR analysis, the preference for $w<-1$ and the anti-/correlation of $w$ with the slope and amplitude parameters of the richness--mass relation shifts the corresponding posteriors into the same region of the parameter space preferred by the full data combination (see figure \ref{fig:wCDM_constraints_all} in appendix \ref{app:wCDM}).
Despite the modest ($\sim 0.5-1.0\sigma$) shift of the $\lambda$-M posteriors observed for the $w$CDM model, the resulting mass-richness relations are consistent within one sigma with the corresponding results of the \LCDM analysis.

Adopting the DIC to asses which cosmological model performs better, we find a "strong" preference for the $w$CDM over the \LCDM model: ${\rm DIC}^{\Lambda{\rm CDM}}-{\rm DIC}^{w{\rm CDM}}=-5.3$ for DES-NC+SPT-OMR and ${\rm DIC}^{\Lambda{\rm CDM}}-{\rm DIC}^{w{\rm CDM}}=-11.3$ for the full data combination. This preference is mainly driven by the improved fit to DES-NC data compared to \LCDM case in all the redshift/richness bins, though a larger than $2\sigma$ tension persist with the highest richness/redshift data point.
Nevertheless, with the current level of knowledge of the scaling relations and their evolution it is not clear if the preference for a $w$CDM is driven by a flawed modeling of the scaling relation absorbed by $w$, or an actual preference for an evolving dark energy cosmology.

Not surprisingly, given the broad posteriors derived for $w$, 
our results for the dark energy equation of state parameter are consistent with those obtained from \planck\ CMB data ($w=-1.41 \pm 0.27$; \textit{green} contours) and DES Y1 galaxy clustering and shear analysis ($w=-0.88^{+0.26}_{-0.15}$; \textit{pink} contours), as well as, with those derived in the SPT-SZ \citep[$w=-1.55\pm 0.41$;][]{Bocquet2018} and WtG \citep[$w=-0.98\pm 0.15$, assuming $\sum m_\nu=0$ and including gas mass fraction data and a $\pm 5$ per cent uniform prior on the redshift evolution of the $M_{\rm gas}$--$M$ relation;][]{mantzetal16} cluster abundance studies.
As mentioned above, an improved calibration of the scaling relations and their evolution will be paramount for future cluster surveys aimed to disentangle a cosmological constant from a $w$CDM model \citep[e.g.][]{Sartoris2016}.

\section{Summary and Conclusion}
\label{sec:concl}
In this study, we derive cosmological and scaling relation constraints from the combination of DES Y1 cluster abundance data (DES-NC) and SPT follow-up data (SPT-OMR). 
The former contains $\sim 6500$ clusters above richness $20$ in the redshift range $0.2<z<0.65$, the latter consists of high-quality X-ray data from \textit{Chandra} and imaging data from HST and Megacam for 121 clusters collected within the SPT-SZ 2500$\deg^2$ survey, along with richness estimates for 129 systems cross matched with the DES Y1 \redmapper\ catalog.
The SPT multi-wavelength data allows us to constrain the richness--mass scaling relation, enabling the cosmological exploitation of the DES cluster counts data.
Mass proxies based on photometric data are prone to contaminations from structures along the line of sight --- i.e. projection effects --- which hamper the calibration of the scaling relations.
To explore possible model systematics related to the latter we consider two calibrations of the observational scatter on richness estimates: i) a simple Gaussian model which accounts only for the noise due to misclassification of background and member galaxies, and ii) the model developed in \cite{costanzietal18} which includes also the scatter on $\lob$ introduced by projection effects (labelled respectively BKG and PRJ throughout the paper).

Independently from the model adopted for the scatter on the observed richness, we derive cosmological constraints for a \LCDM model consistent with CMB data and low redshift probes, including other cluster abundance studies. 
Our results are in contrast with the findings of \citetalias{desy1cl} which obtained cosmological constraints in tension with multiple cosmological probes analysing the same DES abundance data but calibrating the $\lambda-M$ relation with mass estimates derived from stacked weak lensing data. Our results thus support the conclusion of \citetalias{desy1cl} which suggests that the tension is due to the presence of systematics in the modeling of the stacked weak lensing signal of low richness clusters ($\lob \lesssim 30$).
Indeed, the mass--richness relations derived in this work adopting the BKG and PRJ models are in tension with that derived in \citetalias{desy1cl} below $\lob \sim 60$ and $\lob \sim 40$, respectively.
We stress however that the SPT-OMR data are mainly available for $\lob \gtrsim 40$ systems and thus we need to extrapolate the $\lob-M$ relation when fitting the DES abundance data at lower richness.
Nevertheless, both scatter models perform well in fitting the DES cluster abundance at all richnesses, supporting the goodness of the relation extrapolated at low richness.

We further consider the combination of the DES-NC and SPT-OMR data with the SPT number counts data above redshift $z=0.65$ (SPT-NC), to assess possible cosmological gains given by the analysis of the joint abundance catalog. This also serves as a test of the consistency of the three combined data sets.
When included in the analysis the SPT-NC data reduces the $\sigma_8$ and $\Omegam$ uncertainties by $30\%$ and $20\%$ respectively, while shifting their posteriors along the $S_8$ degeneracy direction,  increasing the tension with other cosmological probes,
and especially with the SPT-SZ 2500 results, with which it shares the SPT abundance at $z>0.65$ and follow-up data.
The shift is due to the tension between the scaling relation parameters preferred by the DES and SPT abundance data and the SPT follow-up data at the "fiducial" cosmology $\sigma_8 \sim 0.75$ $\Omegam \sim 0.3$. This tension is largely solved once we consider the PRJ model. Compared to the BKG results, it provides cosmological posteriors for the full data combination in better agreement with all the other probes considered here.
Adopting the DIC for the model selection, we find a "positive" preference for the BKG model for the DES-NC+SPT-OMR data combination, and a "positive" preference for the PRJ model for the full data combination. Additional follow-up data, especially at low richness will be necessary to clearly identify which scatter model for $\lob$ is best suited to describe the data. In this respect, the upcoming SZ and X-ray surveys SPT-3G and eROSITA are expected to provide valuable follow-up data by lowering the limiting mass of the detected clusters to $\sim 10^{14} \msun$ \citep[see e.g.][]{Grandis2019}.

Finally we consider a $w$CDM model and derive cosmological constraints for the DES-NC+SPT-OMR and DES-NC+SPT-[OMR,NC] data combinations assuming the BKG model. We find in both cases a preference at more than 1 $\sigma$ for $w$ values lower than $-1$, but consistent with a cosmological constant.
The inclusion of the SPT-NC does not substantially improve the $w$ constraints despite the larger redshift leverage provided by the SPT abundance data, indicating that also in this case we are limited by the uncertainty in the calibration of the scaling relations and their evolution.
According to the DIC the $w$CDM model is "strongly" preferred over the \LCDM one, thanks to the improved fit to the DES-NC data provided by the extended model. However, given the strong degeneracy between $w$ and the scaling relation parameters we cannot exclude that this preference is due to a flawed modeling of the scaling relations which is absorbed by $w$. Again, an improved calibration of the scaling relations and their evolution, will be necessary for future cluster surveys aimed to constrain the dark energy equation of state parameter. Future optical survey such as Euclid and LSST, in combination with data from the forthcoming eROSITA and SPT-3G surveys, will provide the necessary high-redshift multi-wavelength data to break such degeneracies and thus constrain parameters affecting the growth rate of cosmic structures \citep[see e.g.][]{Sartoris2016}.

The results of this work highlight the capability of multi-wavelength cluster data to improve our understanding of  the systematics affecting the observable-mass scaling relations, and the potential power that a joint analysis of cluster catalogs detected at different wavelengths will have in future cosmological studies with galaxy clusters.


\section*{Acknowledgements}  
\label{sec:acknowledgements}

Funding for the DES Projects has been provided by the U.S. Department of Energy, the U.S. National Science Foundation, the Ministry of Science and Education of Spain, 
the Science and Technology Facilities Council of the United Kingdom, the Higher Education Funding Council for England, the National Center for Supercomputing 
Applications at the University of Illinois at Urbana-Champaign, the Kavli Institute of Cosmological Physics at the University of Chicago, 
the Center for Cosmology and Astro-Particle Physics at the Ohio State University,
the Mitchell Institute for Fundamental Physics and Astronomy at Texas A\&M University, Financiadora de Estudos e Projetos, 
Funda{\c c}{\~a}o Carlos Chagas Filho de Amparo {\`a} Pesquisa do Estado do Rio de Janeiro, Conselho Nacional de Desenvolvimento Cient{\'i}fico e Tecnol{\'o}gico and 
the Minist{\'e}rio da Ci{\^e}ncia, Tecnologia e Inova{\c c}{\~a}o, the Deutsche Forschungsgemeinschaft and the Collaborating Institutions in the Dark Energy Survey. 

The Collaborating Institutions are Argonne National Laboratory, the University of California at Santa Cruz, the University of Cambridge, Centro de Investigaciones Energ{\'e}ticas, 
Medioambientales y Tecnol{\'o}gicas-Madrid, the University of Chicago, University College London, the DES-Brazil Consortium, the University of Edinburgh, 
the Eidgen{\"o}ssische Technische Hochschule (ETH) Z{\"u}rich, 
Fermi National Accelerator Laboratory, the University of Illinois at Urbana-Champaign, the Institut de Ci{\`e}ncies de l'Espai (IEEC/CSIC), 
the Institut de F{\'i}sica d'Altes Energies, Lawrence Berkeley National Laboratory, the Ludwig-Maximilians Universit{\"a}t M{\"u}nchen and the associated Excellence Cluster Universe, 
the University of Michigan, NFS's NOIRLab, the University of Nottingham, The Ohio State University, the University of Pennsylvania, the University of Portsmouth, 
SLAC National Accelerator Laboratory, Stanford University, the University of Sussex, Texas A\&M University, and the OzDES Membership Consortium.

Based in part on observations at Cerro Tololo Inter-American Observatory at NSF's NOIRLab (NOIRLab Prop. ID 2012B-0001; PI: J. Frieman), which is managed by the Association of Universities for Research in Astronomy (AURA) under a cooperative agreement with the National Science Foundation.

The DES data management system is supported by the National Science Foundation under Grant Numbers AST-1138766 and AST-1536171.
The DES participants from Spanish institutions are partially supported by MICINN under grants ESP2017-89838, PGC2018-094773, PGC2018-102021, SEV-2016-0588, SEV-2016-0597, and MDM-2015-0509, some of which include ERDF funds from the European Union. IFAE is partially funded by the CERCA program of the Generalitat de Catalunya.
Research leading to these results has received funding from the European Research
Council under the European Union's Seventh Framework Program (FP7/2007-2013) including ERC grant agreements 240672, 291329, and 306478.
We  acknowledge support from the Brazilian Instituto Nacional de Ci\^encia
e Tecnologia (INCT) do e-Universo (CNPq grant 465376/2014-2).

This manuscript has been authored by Fermi Research Alliance, LLC under Contract No. DE-AC02-07CH11359 with the U.S. Department of Energy, Office of Science, Office of High Energy Physics. This paper has gone through internal review by the DES collaboration.

MC and AS are supported by the ERC-StG 'ClustersXCosmo' grant agreement 716762. AS is supported by the FARE-MIUR grant 'ClustersXEuclid'. AAS acknowledges support by U.S. NSF grant AST-1814719.
This analysis has been carried out using resources of the computing center of INAF-Osservatorio Astronomico di Trieste, under the coordination of the CHIPP project \citep{bertocco2019, taffoni2020}, CINECA grants INA20\_C6B51 and INA17\_C5B32, and of the National Energy Research Scientific Computing Center (NERSC), a U.S. Department of Energy Office of Science User Facility operated under Contract No. DE-AC02-05CH11231.



\bibliography{database.bib} 


\appendix


\section{\LCDM results: $Y_x$ scaling relation and correlation coefficients}
\label{app:yx}
\begin{figure*}
\begin{center}
    \includegraphics[width=\textwidth]{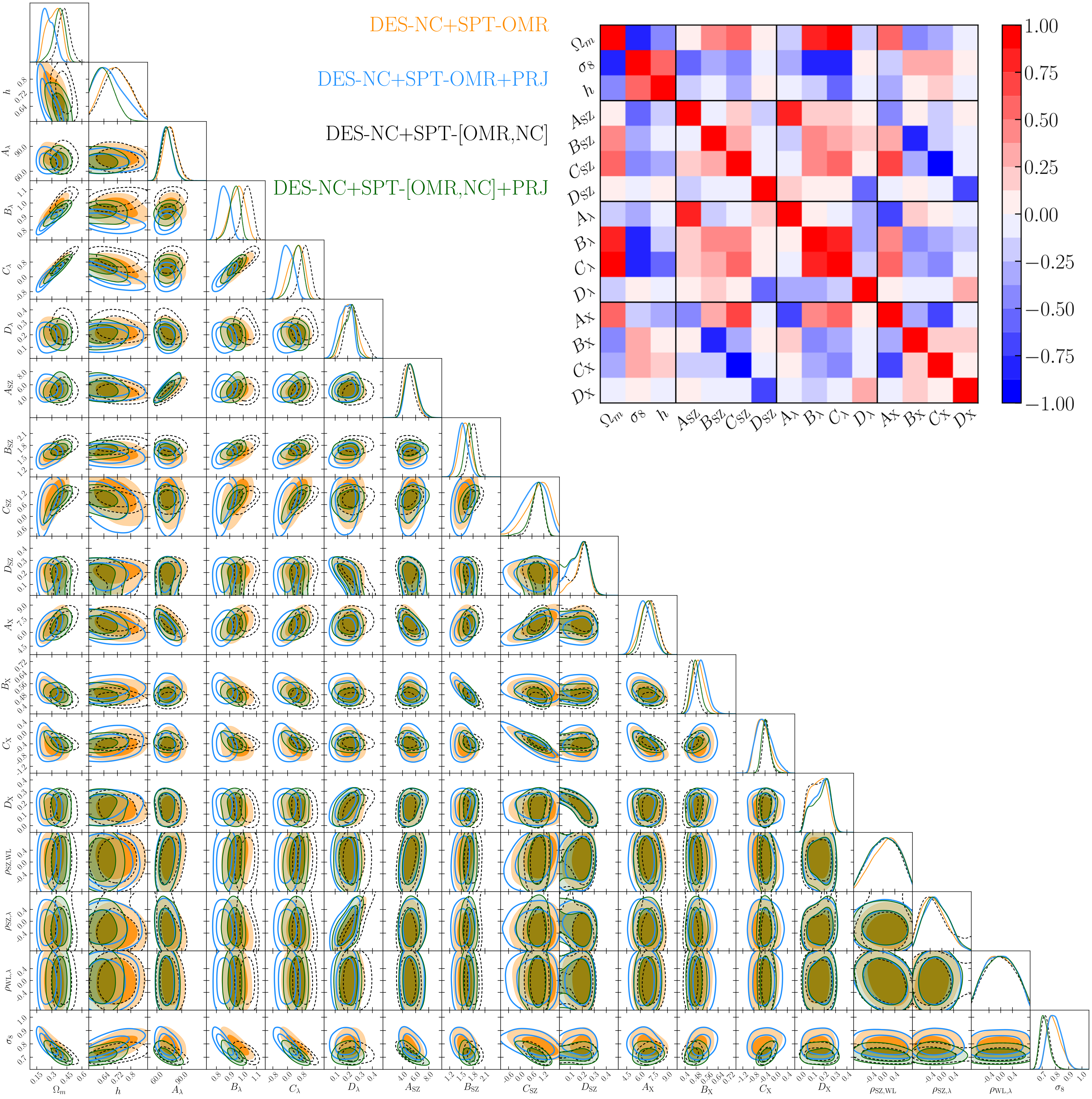}
\end{center}
\caption{Marginalized posterior distributions for the $\Lambda$CDM+$\sum m_\nu$ model for a subset of the fitted parameters. The $2D$ contours correspond to the $68\%$ and $95\%$ confidence levels of the marginalized posterior distribution. The description of the model parameters along with their posteriors are listed in Table \ref{tab:parameters}.  \textit{Inset} panel: Correlation matrix for the scaling relations and cosmological parameters derived from the DES-NC+SPT-OMR analysis. }
\label{fig:LCDM_constraints_all}
\end{figure*}

For completeness we report in figure \ref{fig:LCDM_constraints_all} the posteriors obtained for the \LCDM model including the $Y_x$ scaling relation parameters and the correlation coefficients. 
Also, to easily visualize the many degeneracies between the parameters constrained in the analysis  we show in the inset plot of figure \ref{fig:LCDM_constraints_all} the correlation matrix obtained from the DES-NC+SPT-OMR data. The correlation matrices for the full data combination and/or including the PRJ model are qualitatively consistent with the one shown here.
Depending only on the SPT-OMR data the $Y_x$ posteriors are consistent among the different analyses, even though the correlations with the other scaling relations cause slight shifts of the slope and amplitude parameters and improve the constraint on the evolution parameter once we include the SPT-NC data.

%
%

\section{Observed richness--mass scaling relations}
\label{app:lm}
To ease the comparison and use of our results we report here the mean observed richness--mass scaling relations derived from the DES-NC+SPT-OMR data combination for the two scatter models adopted.
The mean relations and uncertainties are derived from the appropriate model for $P(\lob|M)= \int \de \lambda P(\lob|\lambda,z) P(\lambda|M,z)$ sampling the posterior distributions of the richness--mass relation.
Fitting the mean relation to a power law model we obtain for the BKG model:
\begin{multline}
    \langle \lob |M_{500,c},z \rangle = 79.8\pm 5.0 \left(\frac{M_{500,c}}{3 \times 10^{14} \msunh}\right)^{0.93\pm 0.03} \\ \left(\frac{1+z}{1+0.45}\right)^{-0.49^{+0.71}_{-0.80}}
\end{multline}
while for the PRJ model we obtain:
\begin{multline}
    \langle \lob |M_{500,c},z \rangle = 80.1 \pm 4.1 \left(\frac{M_{500,c}}{3 \times 10^{14} \msunh}\right)^{0.88\pm 0.03} \\ \left(\frac{1+z}{1+0.45}\right)^{-0.06\pm 0.6}
\end{multline}


\section{$w$CDM results: scaling relations and correlation coefficients}
\label{app:wCDM}
\begin{figure*}
\begin{center}
    \includegraphics[width=\textwidth]{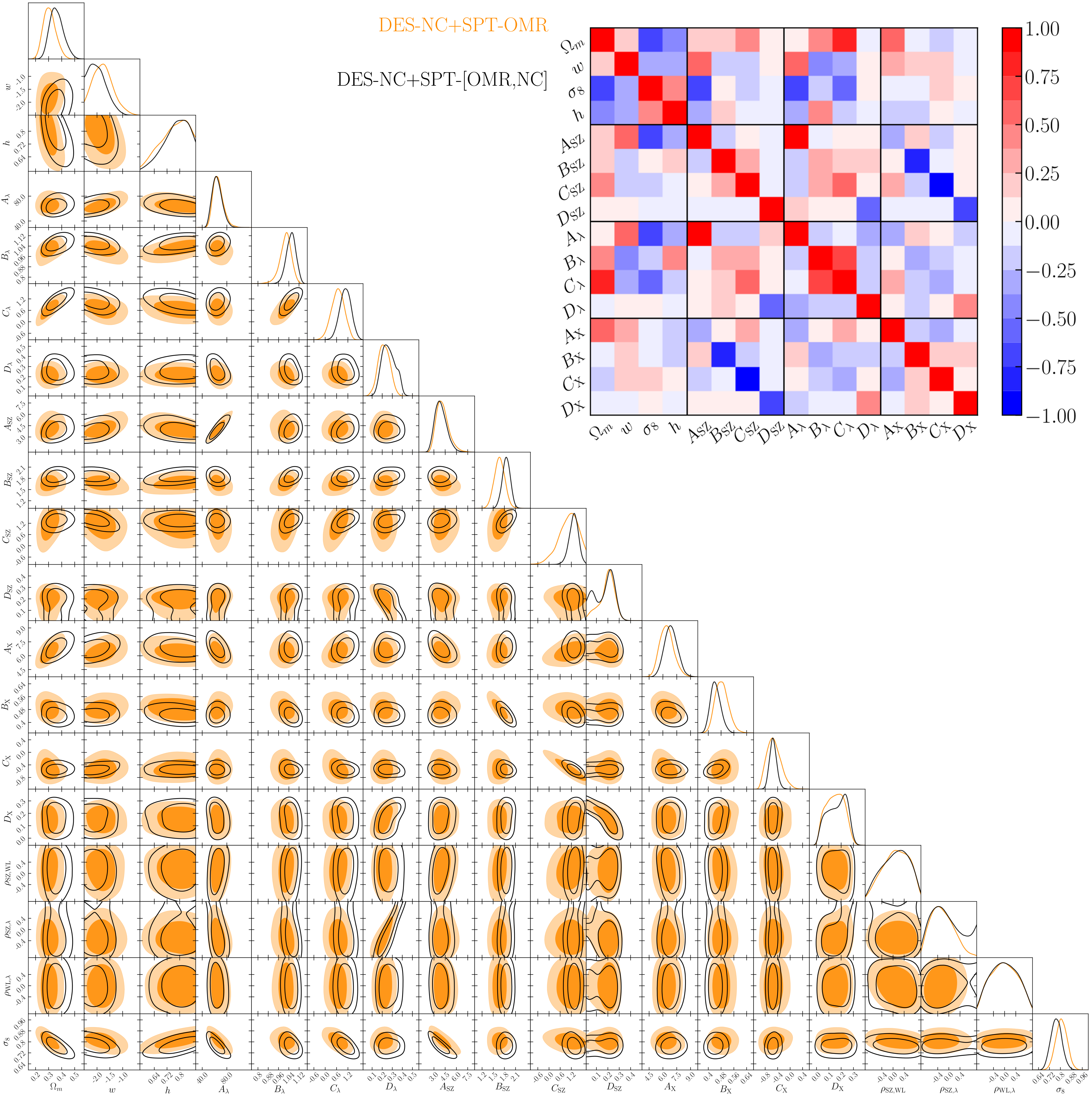}
\end{center}
\caption{Marginalized posterior distributions for the $w$CDM+$\sum m_\nu$ model. The $2D$ contours correspond to the $68\%$ and $95\%$ confidence levels of the marginalized posterior distribution. The description of the model parameters along with their posteriors are listed in Table \ref{tab:parameters}.  \textit{Inset} panel: Correlation matrix for the scaling relations and cosmological parameters derived from the DES-NC+SPT-OMR analysis. }
\label{fig:wCDM_constraints_all}
\end{figure*}

As for the \LCDM analysis we report in figure \ref{fig:wCDM_constraints_all} the posteriors obtained for the $w$CDM model including the scaling relation parameters and the correlation coefficients omitted in the main text. 
The inset plot in figure shows the correlation matrix for a subset of the varied parameters obtained from the DES-NC+SPT-OMR analysis.

%
%

\end{document}